\RequirePackage[l2tabu, orthodox]{nag}
\RequirePackage{snapshot}

\documentclass[10pt,onecolumn]{extarticle}
\makeatletter
\if@twocolumn
  \usepackage[dvips,letterpaper,top=0.5in, bottom=0.5in, left=0.75in, right=0.5in,includefoot,heightrounded]{geometry}
\else
  \usepackage[dvips,letterpaper,margin=1in,includefoot,heightrounded]{geometry}
\fi

\usepackage{srcltx}

\usepackage{amsmath}
\usepackage{amssymb,amsfonts}

\usepackage{abstract}

\usepackage{graphics}
\usepackage{graphicx}
\usepackage[labelformat=simple]{subcaption}

\usepackage[usenames,dvipsnames]{color}
\usepackage[usenames,dvipsnames,table]{xcolor}

\usepackage{booktabs}

\usepackage{setspace}
\usepackage{flushend}
\usepackage{multicol}

\usepackage{cite}
\usepackage{url}
\usepackage[normalem]{ulem}

\usepackage{enumerate}

\usepackage{multirow}
\usepackage{array}

\usepackage{hyperref}

\usepackage{lipsum}

\usepackage[sc,tiny,center]{titlesec}

\usepackage{microtype}

\makeatletter

\newenvironment{rsmallmatrix}{\null\,\vcenter\bgroup
  \Let@\restore@math@cr\default@tag
  \baselineskip6\ex@ \lineskip1.5\ex@ \lineskiplimit\lineskip
  \ialign\bgroup\hfil$\m@th\scriptstyle##$&&\thickspace\hfil
  $\m@th\scriptstyle##$\crcr
}{%
  \crcr\egroup\egroup\,%
}
\makeatother

\newcommand{\printtitle}{%
\makeatletter
\if@twocolumn

\twocolumn[%
  \maketitle
  \begin{onecolabstract}
    \myabstract
  \end{onecolabstract}
  \begin{center}
    \small
    \textbf{Keywords}
    \\\medskip
    \mykeywords
  \end{center}
  \bigskip
]
\saythanks
\else
  \maketitle
  \begin{abstract}
    \myabstract
  \end{abstract}
  \begin{center}
    \small
    \textbf{Keywords}
    \\\medskip
    \mykeywords
  \end{center}
  \bigskip
  \onehalfspacing
\fi
}

\title{%
Energy-efficient 8-point DCT Approximations: Theory and Hardware Architectures
}

\author{%
Renato~J.~Cintra%
\thanks{
Renato~J.~Cintra
is
with the
Signal Processing Group,
Departamento de Estat\'istica,
Universidade Federal de Pernambuco,
Recife, PE, Brazil;
Equipe Cairn, IRISA-INRIA, Universit\'e de Rennes~1, Rennes, France;
LIRIS, Institut National des Sciences Appliqu\'ees, Lyon, France
(\mbox{e-mail:~rjdsc@stat.ufpe.org}).
}
\and
F\'abio~M.~Bayer%
\thanks{%
F\'abio~M.~Bayer
is with the
Departamento de Estat\'istica
and LACESM,
Universidade Federal de Santa Maria,
Santa Maria, RS, Brazil
(e-mail: bayer@ufsm.br).
}
\and
Vitor~A.~Coutinho
\thanks{
Vitor~A.~Coutinho
is
with the
Graduate Program in Electrical Engineering
and
the
Signal Processing Group,
Departamento de Estat\'istica,
Universidade Federal de Pernambuco, Recife, PE, Brazil
(\mbox{e-mail:~vitor.andrade.coutinho@gmail.com}).
}
\and
Sunera~Kulasekera
\and
Arjuna~Madanayake%
\thanks{%
Sunera Kulasekera
and
Arjuna Madanayake
are with the
Department of Electrical and Computer Engineering,
The University of Akron, Akron, OH, USA
(e-mail: arjuna@uakron.edu).}
}

\date{}

\newcommand{\myabstract}{%
Due to
its remarkable energy compaction properties,
the discrete cosine transform (DCT)
is employed in a multitude of compression standards,
such as JPEG and H.265/HEVC.
Several low-complexity integer approximations for the DCT
have been proposed
for both \mbox{1-D}
and \mbox{2-D} signal analysis.
The increasing demand for
low-complexity, energy efficient
methods
require algorithms with even lower computational costs.
In this paper,
new 8-point DCT approximations
with very low arithmetic complexity are presented.
The new transforms are proposed
based on pruning
state-of-the-art DCT approximations.
The proposed algorithms were assessed in terms of
arithmetic complexity,
energy retention capability,
and
image compression performance.
In addition,
a metric combining performance and computational complexity measures was proposed.
Results showed good performance and extremely low computational complexity.
Introduced algorithms were
mapped into systolic-array digital architectures and physically
realized as digital prototype circuits using FPGA technology and
mapped to 45\,nm CMOS technology.
All hardware-related metrics
showed low resource consumption of the proposed pruned approximate transforms.
The best proposed transform according to the introduced metric presents a
reduction in power consumption of 21--25\%.
}

\newcommand{\mykeywords}{%
DCT approximation \and image compression\and FPGA \and pruned transforms
}

\begin{document}

\printtitle

\section{Introduction}

Transform-based methods are widely
employed in digital signal processing applications~\cite{ahmed1975}.
In this context,
the efficient computation of
discrete transforms
has constantly attracted community efforts
and
the proposition of fast algorithms~\cite{Blahut2010}.
In particular,
the 8-point discrete cosine transform (DCT)
has a proven record of scientific and industrial
applications,
as demonstrated by the multitude of
image and video coding standards that adopt it,
such as:
JPEG~\cite{Wallace1992},
MPEG~\cite{Gall1992, roma2007hybrid, mpeg2},
H.261~\cite{h261, Liou1990},
H.263~\cite{h263, roma2007hybrid},
H.264/AVC~\cite{wiegand2003, h264},
and
the recent high efficiency video coding (HEVC)~\cite{hevc,hevc1}.
The HEVC is capable of achieving high compression performance
at
approximately half the bit rate required by its predecessor
H.264/AVC with same image quality~\cite{hevc1, Park2012, Ohm2012, Potluri2013}.
On the other hand,
the HEVC requires a significantly higher
computational
complexity in terms of arithmetic operations~\cite{Park2012, Ohm2012, sullivan2012, Potluri2013},
being 2--4 times more computationally costly
than H.264/AVC~\cite{Park2012, Potluri2013}.
In this context,
the efficient computation of the DCT
is a venue for
improving the performance
of above-mentioned
codecs.

Since its inception,
several fast algorithms for the DCT have been proposed~\cite{Chen1977, hou1987fast, Arai1988, Loeffler1989, fw1992, britanak2007discrete}.
However,
traditional algorithms
aim at the computation of the \emph{exact} DCT,
which requires several multiplication operations.
Additionally,
several algorithms have achieved
theoretical multiplicative complexity lower-bounds~\cite{Loeffler1989,winograd1980}.
As a consequence,
the progress in this area
headed to approximate methods~\cite{Haweel2001,lengwehasatit2004scalable,bas2008}.
In some applications,
a simple DCT approximation
can provide meaningful results at low arithmetic complexity~\cite{cb2011}.
Thus,
approximation techniques for the DCT
are becoming increasingly
popular~\cite{Haweel2001,bas2008, bas2009, bas2013,Cintra2014-sigpro}.
Such approximations can
reduce the computational demands of the DCT,
leading to low-power,
high-speed realizations~\cite{Potluri2013},
while ensuring adequate numerical accuracy.

Furthermore,
it is a well-known fact
that
in many DCT applications~\cite{Makkaoui2010,Docef2002,rao1990discrete},
the most useful signal information tends to be concentrated
in the low-frequency coefficients.
This is because the DCT
presents good energy compaction properties,
which are closely related to the Karhunen-Lo\`eve transform~\cite{Ahmed1974}.
Therefore,
only the low-frequency DCT components
are necessary to be computed in these applications.
A typical example of this situation
occurs in data compression applications~\cite{Rao2001},
where
high-frequency components are often
zeroed by the quantization process~\cite[p.~586]{Malepati2010}.
Then,
only the quantities
that are likely to be significant
should be computed~\cite{Huang2000}.
This approach is called frequency-domain
\emph{pruning}
and has been employed
for computing the discrete Fourier
transform~(DFT)~\cite{wang2012generic, airoldi2010energy, whatmough2012vlsi,kim2011islanding, carugati2012variable}.
Such methodology was originally
applied in the DCT context in~\cite{wang1991pruning}
and~\cite{skodras1994fast}.
In~\cite{Makkaoui2010, Lecuire2012},
the two-dimensional (\mbox{2-D}) version of the pruned
DCT was proposed.
In the context
of low-powered wireless vision sensor networks,
a pruned DCT was proposed in~\cite{kouadria2013low}
based on the binary DCT~\cite{bas2013}.

In~\cite{meher2014efficient},
Meher~\emph{et al.}
proposed a HECV architecture
where the wordlength was maintained fixed
by means of discarding least significant bits.
In that context,
the goal was the minimization of
the computation complexity
at the expense of wordlength truncation.
Such approach was also termed `pruning'.
However,
it is fundamentally different from the approach discussed
in the current paper.
This terminology distinction is worth observing.

Thus,
in response to the growing need
for high compression of image and moving pictures
for various applications~\cite{hevc},
we propose
a further reduction
of the computational cost
of the 8-point DCT computation
in the context of JPEG-like compression
and HEVC processing.
In this work,
we introduce
pruned DCT approximations
for image and video compression.
Essentially,
DCT-like pruning consists of extracting from
a given approximate DCT matrix
a submatrix that aims at furnishing
similar mathematical properties.
We advance the application
of pruning techniques
to several DCT approximations
listed in recent literature.
In this paper,
we aim
at identifying
adequate pruned approximations
for image %
compression applications.
VLSI realizations
of both \mbox{1-D} and \mbox{2-D}
of the proposed methods
are also sought.
This paper is organized as follows.
In Section~\ref{sec:math_back},
a mathematical review
of DCT approximation and pruning methods
is furnished.
Exact and approximate DCT are presented and
the pruning procedure is mathematically described.
In Section~\ref{sec:complex_perform},
we propose
several pruned methods
for approximate DCT computation
and assess them
by means
of
arithmetic complexity,
coefficient energy distribution in transform-domain,
and
image compression performance.
A combined figure of merit
considering performance and complexity
is introduced.
In Section \ref{sec:vlsi},
a VLSI realization of the optimum pruned method
according to the suggested figure of merit
is proposed.
Both FPGA and ASIC realizations
are assessed in terms of
area, time, frequency,
and power consumption.
Section \ref{sec:conclusion}
concludes the paper.

\section{Mathematical Background}
\label{sec:math_back}

\subsection{Discrete Cosine Transform}

Let
$\mathbf{x}=\begin{bmatrix} x_0 & x_1 & \cdots & x_{N-1} \end{bmatrix}^\top$
be an $N$-point input vector.
The one-dimensional
DCT is a linear transformation
that maps $\mathbf{x}$ into
an output
vector~$\mathbf{X}=\begin{bmatrix} X_0 & X_1 & \cdots & X_{N-1} \end{bmatrix}^\top$
of transform coefficients,
according to the following expression~\cite{Oppenheim2010}:
\begin{align}
\label{equation-dct-summation}
X_k
=
\alpha_k
\cdot
\sqrt{\frac{2}{N}}
\cdot
\sum_{n=0}^{N-1}
x_n
\cdot
\cos\left\{\frac{(n+\frac{1}{2})k\pi}{N} \right\}
,
\end{align}
where
$k = 0,1,\ldots,N-1$,
$\alpha_0 = 1/\sqrt{2}$ and
$\alpha_k = 1$, for $k>0$.
In matrix formalism,
\eqref{equation-dct-summation}
is given by:
\begin{align}
\mathbf{X}
=
\mathbf{C}
\cdot
\mathbf{x}
,
\end{align}
where
$\mathbf{C}$ is
the $N$-point DCT matrix
whose entries are expressed according
$c_{m,n}
=
\alpha_m \cdot \sqrt{2/N}\cdot \cos\left\{(n+\frac{1}{2})m\pi/N \right\}$,
$m,n = 0,1,\ldots,N-1$~\cite{britanak2007discrete}.
Being an orthogonal transform,
the inverse transformation
is given by:
$\mathbf{x} = \mathbf{C}^\top \cdot \mathbf{X}$.
Because
DCT
satisfies the kernel separability property,
the
\mbox{2-D} DCT
can be expressed in terms of the \mbox{1-D} DCT.
Let $\mathbf{A}$ be an $N\times N$ matrix.
The forward \mbox{2-D} DCT
operation applied to
$\mathbf{A}$
yields a transform-domain image~$\mathbf{B}$
furnished by:
$\mathbf{B}
=
\mathbf{C}
\cdot
\mathbf{A}
\cdot
\mathbf{C}^\top
$.
In fact,
the \mbox{2-D} DCT
can be computed after
eight column-wise calls of the \mbox{1-D} DCT to~$\mathbf{A}$;
then the resulting intermediate image is
submitted to
eight row-wise calls of the \mbox{1-D} DCT.
In this paper,
we devote our attention to the case $N=8$.

\subsection{DCT Approximations}

In general terms,
a DCT approximation~$\hat{\mathbf{C}}$ is constituted of
the product a low-complexity matrix~$\mathbf{T}$
and a scaling diagonal matrix~$\mathbf{S}$
that ensures orthogonality
or
quasi-orthogonality~\cite{Cintra2014-sigpro}.
Thus, we have $\hat{\mathbf{C}} = \mathbf{S} \cdot \mathbf{T}$~\cite{bas2008,cb2011,bc2012,Potluri2013}.
The entries of the low-complexity matrix
are
defined over the set $\{0\,\pm1,\pm2\}$,
which results in a multiplierless operator---only
addition and bit-shifting operations are required.
Usually possessing irrational elements,
the scaling diagonal matrix~$\mathbf{S}$ does not
pose any extra computation overhead
for image and video compression applications.
This is due to the fact that the matrix~$\mathbf{S}$
can be conveniently merged into the quantization step
of compression algorithms~\cite{bas2008,bas2009,bc2012,Potluri2013}.

Among the various DCT approximations archived in literature,
we separate the following methods:
(i)~the signed DCT~(SDCT),
which is the seminal method in the DCT approximation field~\cite{Haweel2001};
(ii)~Bouguezel-Ahmad-Swamy approximations~\cite{bas2008,bas2009,bas2013};
(iii)~the rounded DCT (RDCT)~\cite{cb2011},
and
(iv)~the modified RDCT (MRDCT)~\cite{bc2012}.
These approximations were selected because they
collectively exhibit
a wide range of
complexity vs. performance trade-off figures~\cite{bc2012}.
Moreover,
such approximations
have been demonstrated
to be useful in
image compression.
The low-complexity matrices of
above methods are shown in Table~\ref{table-approximations}.
Additionally,
we also
considered
the 8-point naturally ordered Walsh-Hadamard transform~(WHT),
which is a well-known low-complexity transform
with applications in image processing~\cite{Elliot1982,bas2013}.

\begin{table}
\centering
\caption{Approximate DCT methods}
\label{table-approximations}
\begin{tabular}{lcc}
\toprule
Method &
$\mathbf{T}$ &
Orthogonal?
\\
\midrule
SDCT~\cite{Haweel2001} &
$
\left[
\begin{rsmallmatrix}
 1& 1& 1& 1& 1& 1& 1& 1 \\
 1& 1& 1& 1&-1&-1&-1&-1 \\
 1& 1&-1&-1&-1&-1& 1& 1 \\
 1&-1&-1&-1& 1& 1& 1&-1 \\
 1&-1&-1& 1& 1&-1&-1& 1 \\
 1&-1& 1& 1&-1&-1& 1&-1 \\
 1&-1& 1&-1&-1& 1&-1& 1 \\
 1&-1& 1&-1& 1&-1& 1&-1
\end{rsmallmatrix}
\right]
$
&
No
\\
WHT~\cite{Elliot1982}
&
$
\left[
\begin{rsmallmatrix}
     1   & 1  &  1  &  1  &  1  &  1  &  1  &  1\\
     1  & -1  &  1  & -1  &  1  & -1  &  1  & -1\\
     1  &  1  & -1  & -1  &  1  &  1  & -1  & -1\\
     1  & -1  & -1  &  1  &  1  & -1  & -1  &  1\\
     1  &  1  &  1  &  1  & -1  & -1  & -1  & -1\\
     1  & -1  &  1  & -1  & -1  &  1  & -1  &  1\\
     1  &  1  & -1  & -1  & -1  & -1  &  1  &  1\\
     1  & -1 &  -1  &  1  & -1  &  1  &  1  & -1
\end{rsmallmatrix}
\right]
$
&
Yes
\\
BAS-2008~\cite{bas2008} &
$
\left[
\begin{rsmallmatrix}
 1& 1& 1& 1& 1& 1& 1& 1 \\
 1& 1& 0& 0& 0& 0&-1&-1 \\
 1& \frac{1}{2}&-\frac{1}{2}&-1&-1&-\frac{1}{2}& \frac{1}{2}& 1 \\
 0& 0&-1& 0& 0& 1& 0& 0 \\
 1&-1&-1& 1& 1&-1&-1& 1 \\
 1&-1& 0& 0& 0& 0& 1&-1 \\
 \frac{1}{2}&-1& 1&-\frac{1}{2}&-\frac{1}{2}& 1&-1& \frac{1}{2} \\
 0& 0& 0&-1& 1& 0& 0& 0
\end{rsmallmatrix}
\right]
$
&
Yes
\\
BAS-2009~\cite{bas2009} &
$
\left[
\begin{rsmallmatrix}
 1& 1& 1& 1& 1& 1& 1& 1 \\
 1& 1& 0& 0& 0& 0&-1&-1 \\
 1& 1&-1&-1&-1&-1& 1& 1 \\
 0& 0&-1& 0& 0& 1& 0& 0 \\
 1&-1&-1& 1& 1&-1&-1& 1 \\
 1&-1& 0& 0& 0& 0& 1&-1 \\
 1&-1& 1&-1&-1& 1&-1& 1 \\
 0& 0& 0&-1& 1& 0& 0& 0
\end{rsmallmatrix}
\right]
$
&
Yes
\\
BAS-2013~\cite{bas2013} &
$
\left[
\begin{rsmallmatrix}
 1& 1& 1& 1& 1& 1& 1& 1 \\
 1& 1& 1& 1&-1&-1&-1&-1 \\
 1& 1&-1&-1&-1&-1& 1& 1 \\
 1& 1&-1&-1& 1& 1&-1&-1 \\
 1&-1&-1& 1& 1&-1&-1& 1 \\
 1&-1&-1& 1&-1& 1& 1&-1 \\
 1&-1& 1&-1&-1& 1&-1& 1 \\
 1&-1& 1&-1& 1&-1& 1&-1
\end{rsmallmatrix}
\right]
$
&
Yes
\\
RDCT~\cite{cb2011} &
$
\left[
\begin{rsmallmatrix}
1 & 1 & 1 & 1 & 1 & 1 & 1 & 1 \\
1 & 1 & 1 & 0 & 0 &-1 &-1 &-1 \\
1 & 0 & 0 &-1 &-1 & 0 & 0 & 1 \\
1 & 0 &-1 &-1 & 1 & 1 & 0 &-1 	\\
1 &-1 &-1 & 1 & 1 &-1 &-1 & 1 \\
1 &-1 & 0 & 1 &-1 & 0 & 1 &-1 \\
0 &-1 & 1 & 0 & 0 & 1 &-1 & 0 \\
0 &-1 & 1 &-1 & 1 &-1 & 1 &  0
\end{rsmallmatrix}
\right]
$
&
Yes
\\
MRDCT~\cite{bc2012} &
$
\left[
\begin{rsmallmatrix}
1 & 1 & 1 & 1 & 1 & 1 & 1 & 1 \\
1 & 0 & 0 & 0 & 0 & 0 & 0 &-1 \\
1 & 0 & 0 &-1 &-1 & 0 & 0 & 1 \\
0 & 0 &-1 & 0 & 0 & 1 & 0 & 0 \\
1 &-1 &-1 & 1 & 1 &-1 &-1 & 1 \\
0 &-1 & 0 & 0 & 0 & 0 & 1 & 0 \\
0 &-1 & 1 & 0 & 0 & 1 &-1 & 0 \\
0 & 0 & 0 &-1 & 1 & 0 & 0 & 0
\end{rsmallmatrix}
\right]
$
&
Yes
\\
\bottomrule
\end{tabular}
\end{table}

\subsection{Pruned Exact and Approximate DCT}

Essentially,
DCT pruning consists of extracting from
the 8$\times$8 DCT matrix~$\mathbf{C}$
a submatrix that aims at furnishing
similar mathematical properties as~$\mathbf{C}$.
Pruning is often realized on the transform-domain
by means of computing fewer transform coefficients
than prescribed by the full transformation.
Usually,
only the $K<N$ coefficients that retain more energy are preserved.
For the DCT,
this corresponds to the first~$K$ rows
of the DCT matrix.
Therefore,
this particular type of pruning
implies the following
$K\times 8$
matrix:
\begin{align}
\mathbf{C}_K
=
\begin{bmatrix}
c_{0,0} & c_{0,1} & \cdots  & c_{0,7} \\
c_{1,0} & c_{1,1} & \cdots  & c_{1,7} \\
\vdots & \vdots & \ddots  & \vdots     \\
c_{K-1,0} & c_{K-1,1} & \cdots  & c_{K-1,7}
\end{bmatrix}
,
\end{align}
where
$0<K\le8$ and
$c_{m,n}$, $m,n=0,1,\ldots,7,$
are the entries of~$\mathbf{C}$.
The case $K=8$ corresponds to the original transformation.
Such procedure was proposed in~\cite{Makkaoui2010,Lecuire2012}
for the DCT in the context of wireless sensor networks.
For the \mbox{2-D}
case,
we have that the pruned DCT is given by:
$\tilde{\mathbf{B}} = \mathbf{C}_K \cdot
\mathbf{A} \cdot \mathbf{C}_K^\top$.
Notice that $\tilde{\mathbf{B}}$ is a $K\times K$ matrix
over the transform-domain.
Lecuire~\emph{et al.}~\cite{Lecuire2012}
showed that retaining the transform-domain coefficients
in a $K\times K$ square pattern
at the upper-right corner
leads to a better energy-distortion trade-off
when compared to the alternative triangle pattern~\cite{Makkaoui2010}.

The pruning approach can be applied to DCT approximations.
By discarding the lower rows of the low-complexity
matrix~$\mathbf{T}$,
we obtain the
following
$K\times N$ pruned matrix transformation:
\begin{align}
\mathbf{T}_K
=
\begin{bmatrix}
t_{0,0} & t_{0,1} & \cdots  & t_{0,7} \\
t_{1,0} & t_{1,1} & \cdots  & t_{1,7} \\
\vdots  & \vdots  & \ddots  & \vdots     \\
t_{K-1,0} & t_{K-1,1} & \cdots  & t_{K-1,7}
\end{bmatrix}
,
\end{align}
where
$t_{m,n}$, $m,n=0,1,\ldots,7$,
are the entries of~$\mathbf{T}$
(cf.~Table~\ref{table-approximations}).
Considering the orthogonalization method
described in~\cite{Cintra2014-sigpro},
the $K\times$8 pruned approximate DCT
is given by:
\begin{align}
\hat{\mathbf{C}}_K
=
\mathbf{S}_K
\cdot
\mathbf{T}_K
,
\end{align}
where
$\mathbf{S}_K = \sqrt{ \operatorname{diag}\{(\mathbf{T}_K \cdot \mathbf{T}_K^\top)^{-1}\}}$
is a $K\times K$ diagonal matrix
and
$\operatorname{diag}(\cdot)$
returns a diagonal matrix with the diagonal elements of its argument.
If $\mathbf{T}$ is orthogonal,
then $\mathbf{T}_K$ satisfies semi-orthogonality~\cite[p.~84]{Abadir2005}.

The \mbox{2-D} pruned DCT of a matrix $\mathbf{A}$ is given by
\begin{align}
\label{equation-pruned-2d}
\tilde{\mathbf{B}} = \mathbf{T}_K \cdot
\mathbf{A} \cdot \mathbf{T}_K^\top
.
\end{align}
Resulting transform-domain matrix $\tilde{\mathbf{B}}$ is sized $K\times K$.

\section{Complexity and Performance Assessment}
\label{sec:complex_perform}

In this section,
we analyze the arithmetic complexity
of the selected pruned DCT approximations.
We also assess
their performance in terms of
energy retention
and
image compression
for each value of~$K$.

\subsection{Arithmetic complexity}

\begin{figure*}%
\centering
 \begin{subfigure}[b]{0.40\linewidth}
        \includegraphics[scale=1.0]{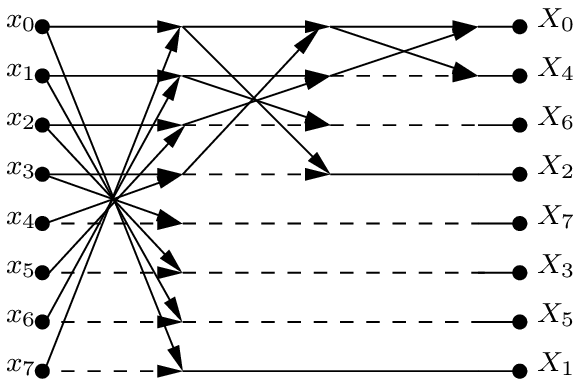}
        \caption{Original MRDCT (14 additions)}
        \label{figure-sfg-mrdct_full}
 \end{subfigure}
 \begin{subfigure}[b]{0.40\linewidth}
 		\includegraphics[scale=1.00]{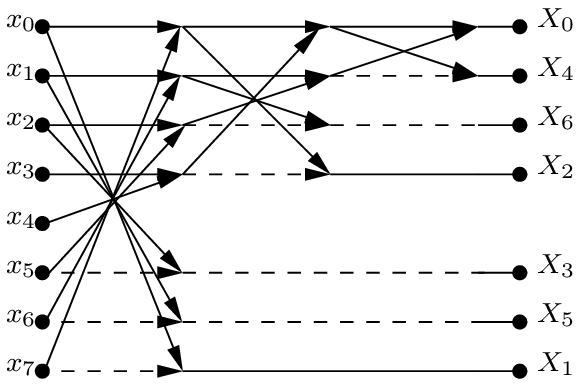}
        \caption{$K = 7$ (13 additions)}
		\label{figure-pruned-mrdct-k7}
 \end{subfigure}
 \begin{subfigure}[b]{0.40\linewidth}
 		\includegraphics[scale=1.00]{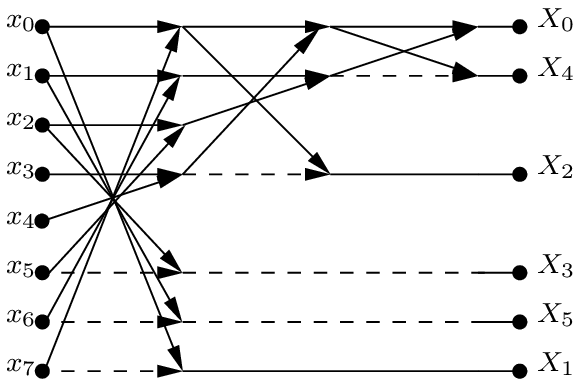}
        \caption{$K = 6$ (12 additions)}
		\label{figure-pruned-mrdct-k6}
 \end{subfigure}
 \begin{subfigure}[b]{0.40\linewidth}
 		\includegraphics[scale=1.00]{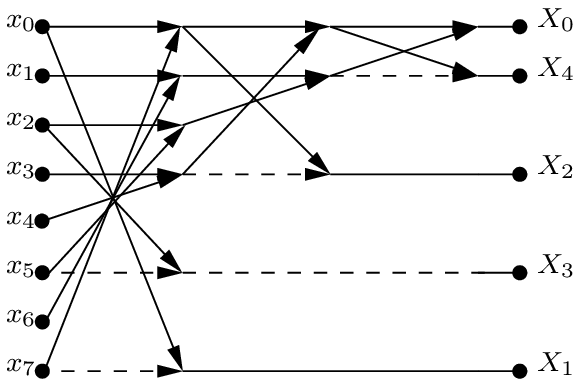}
        \caption{$K = 5$ (11 additions)}
		\label{figure-pruned-mrdct-k5}
\end{subfigure}
 \begin{subfigure}[b]{0.40\linewidth}
 		\includegraphics[scale=1.00]{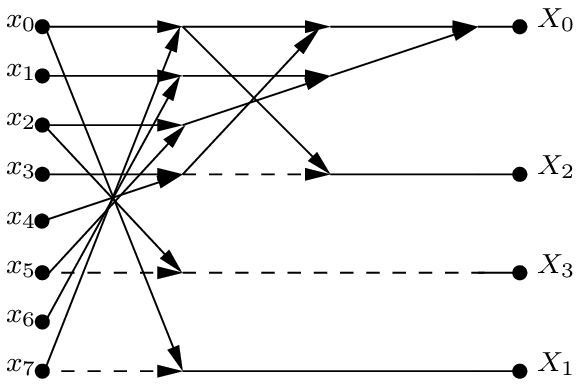}
        \caption{$K = 4$ (10 additions)}
		\label{figure-pruned-mrdct-k4}
 \end{subfigure}
 \begin{subfigure}[b]{0.40\linewidth}
 		\includegraphics[scale=1.00]{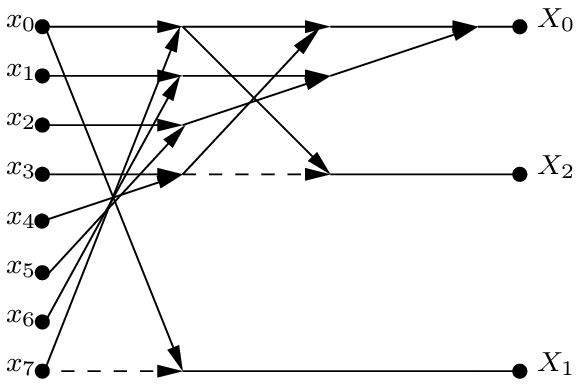}
        \caption{$K = 3$ (9 additions)}
		\label{figure-pruned-mrdct-k3}
 \end{subfigure}
  \begin{subfigure}[b]{0.40\linewidth}
 		\includegraphics[scale=1.00]{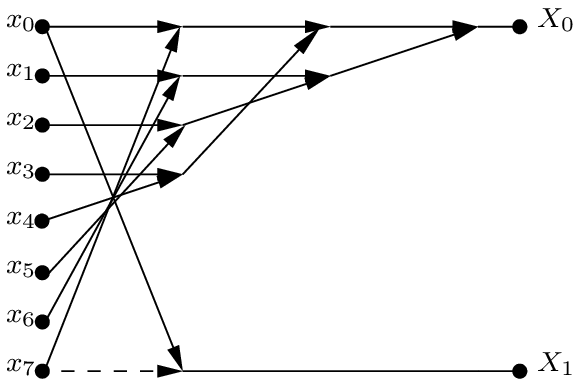}
        \caption{$K = 2$ (8 additions)}
		\label{figure-pruned-mrdct-k2}
\end{subfigure}
\begin{subfigure}[b]{0.40\linewidth}
 		\includegraphics[scale=1.00]{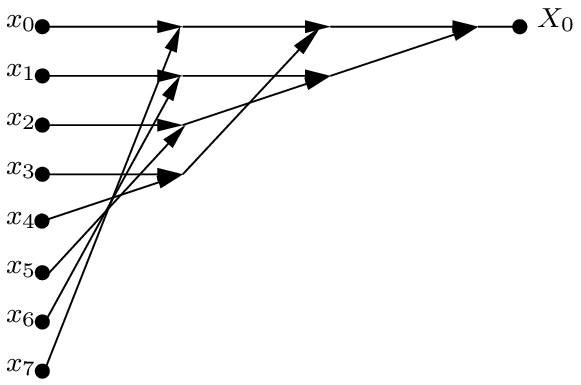}
        \caption{$K = 1$ (7 additions)}
		\label{figure-pruned-mrdct-k1}
\end{subfigure}
\caption{Signal flow graph for the MRDCT matrix and pruned MRDCT matrices}
\label{figure-sfg-mrdct}
\end{figure*}

Because all considered approximate DCT are natively multiplierless operators,
the pruned DCT approximation
inherits such property.
Therefore,
the arithmetic complexity of the pruned approximations
is simply given by the number of additions and bit-shifting operations
required by their respective fast algorithms.
To illustrate the complexity assessment,
we focus on the MRDCT~\cite{bc2012},
whose fast algorithm signal flow graph (SFG)
is shown in \figurename~\ref{figure-sfg-mrdct_full}.
The full computation of the MRDCT requires 14~additions.
By judiciously considering
the computational cost of only the first $K$
transform-domain components,
we derived
fast algorithms for the pruned MRDCT matrices
as
shown in \figurename~\ref{figure-sfg-mrdct}.

The same procedure was applied to each of the discussed approximations
based on their fast algorithms~\cite{Haweel2001, Elliot1982, bas2008, bas2009, bas2013, cb2011, bc2012}.
The obtained arithmetic additive complexity is presented in \tablename~\ref{table-assessment}.
We notice that the pruned MRDCT
exhibited the lowest computational complexity for all values of $K$.
Such mathematical properties of the MRDCT are translated into good hardware designs.
Indeed,
in~\cite{Potluri2013},
several DCT approximations were physically realized in FPGA devices.
Hardware and performance assessments revealed that
the MRDCT outperformed several competitors,
including BAS 2008~\cite{bas2008} and RDCT~\cite{cb2011},
in terms of speed,
hardware resource consumption,
and
power consumption~\cite{Potluri2013}.

An examination of~\eqref{equation-pruned-2d}
reveals
that
the \mbox{2-D} pruned approximate DCT
is computed
after~eight column-wise calls of the
\mbox{1-D} pruned approximate DCT
and~$K$ row-wise call of \mbox{1-D} pruned approximate DCT.
Let $\operatorname{A}_\text{1-D}(\mathbf{T}_K)$
be the additive complexity
of $\mathbf{T}_K$.
Therefore,
the additive complexity of the
\mbox{2-D} pruned approximate DCT is given by:
\begin{equation}
\begin{split}
\operatorname{A}_\text{2-D}(\mathbf{T}_K)
&
=
8 \cdot \operatorname{A}_\text{1-D}(\mathbf{T}_K)
+
K \cdot \operatorname{A}_\text{1-D}(\mathbf{T}_K)
\\
&
=
(8+K) \cdot \operatorname{A}_\text{1-D}(\mathbf{T}_K)
.
\end{split}
\end{equation}

For the particular case of the pruned MRDCT,
we can derive the expressions below:
\begin{align}
\operatorname{A}_\text{1-D}(\mathbf{T}_K)
&
=
K + 6
,
\\
\operatorname{A}_\text{2-D}(\mathbf{T}_K)
&
=
K^2 + 14 \cdot K + 48
,
\end{align}
for
$K = 1,2,\ldots,8$.

\begin{table*}
\centering
\caption{Complexity and performance assessment of %
pruned DCT approximations}
\label{table-assessment}
\begin{tabular}{l l  c c c c c c c c }
\toprule
\multirow{3}{*}{Measure} & \multirow{3}{*}{Method} & \multicolumn{8}{c}{$K$} \\
\cmidrule{3-10}
&  & 1 & 2 & 3 & 4 & 5 & 6 & 7 & 8 \\
\midrule
& Exact DCT~\cite{Makkaoui2010} & 7 & 20  & 23  & 24  & 25  & 26  & 28  & 29 \\
& WHT~\cite{Elliot1982}&
7 & 8 & 11 & 12 & 19 & 20 & 23 & 24 \\
& SDCT~\cite{Haweel2001} &
7 & 14 & 17 & 19 & 20 & 22 & 23 & 24 \\
Additive & BAS-2008~\cite{bas2008} &
7 & 10 & 13 & 14 & 15 & 16 & 17 & 18 \\
complexity & BAS-2009~\cite{bas2009} &
7 & 10 & 13 & 14 & 15 & 16 & 17 & 18 \\
& BAS-2013~\cite{bas2013,kouadria2013low} &
7 & 14 & 17 & 20 & 21 & 22 & 23 & 24 \\
& RDCT~\cite{cb2011} &
7 & 12 & 13 & 16 & 17 & 19 & 20 & 22 \\
& MRDCT~\cite{bc2012} &
7 & 8 & 9 & 10 & 11 & 12 & 13 & 14 \\
\midrule
& Exact DCT & 95.46  & 97.47  & 98.55  & 99.13  & 99.49  & 99.71  & 99.87  & 100.00  \\
& WHT &95.46  &  95.57  &  96.03  &  96.25  &  98.24  &  98.52  &  99.63  & 100.00    \\
Mean& SDCT & 95.46  & 96.39  & 97.30  & 98.16  & 98.52  & 99.26  & 99.61  & 100.00  \\
retained & BAS-2008 & 95.46  & 97.08  & 98.10  & 98.86  & 99.20  & 99.51  & 99.68  & 100.00  \\
energy & BAS-2009 & 95.46  & 97.08  & 97.96  & 98.71  & 99.04  & 99.35  & 99.68  & 100.00  \\
 & BAS-2013 & 95.46  & 97.18  & 98.08  & 98.76  & 99.10  & 99.44  & 99.77  & 100.00  \\
& RDCT & 95.46  & 97.36  & 98.28  & 98.81  & 99.16  & 99.41  & 99.75  & 100.00  \\
& MRDCT & 95.46  & 96.41  & 97.22  & 97.91  & 98.22  & 99.34  & 99.68  & 100.00  \\
\midrule
& Exact DCT & 23.17 & 26.08 & 28.52 & 30.40 & 31.71 & 32.39 & 32.78 & 33.12 \\
& WHT & 23.17  &  23.17  &  23.63  &  23.81  &  26.88  &  27.22  &  29.40  &  30.17   \\
& SDCT & 23.17 & 24.28 & 25.23 & 27.15 & 27.59 & 28.43 & 28.82 & 29.84 \\
Mean & BAS-2008 & 23.17 & 25.30 & 27.04 & 29.34 & 30.15 & 30.97 & 31.33 & 32.20 \\
PSNR & BAS-2009 & 23.17 & 25.30 & 26.95 & 28.70 & 29.47 & 30.14 & 30.96 & 31.76 \\
& BAS-2013 & 23.17 & 24.41 & 26.95 & 28.73 & 29.51 & 30.31 & 31.12 & 31.84 \\
& RDCT & 23.17 & 25.83 & 27.64 & 28.94 & 29.79 & 30.41 & 31.21 & 31.96 \\
& MRDCT & 23.17 & 24.29 & 25.26 & 26.37 & 26.77 & 29.58 & 30.29 & 30.98 \\
\midrule
& Exact DCT & 0.48 & 0.66 & 0.79 & 0.86 & 0.89 & 0.90 & 0.90 & 0.90 \\
& WHT & 0.48  &   0.49  &   0.55  &   0.58  &   0.74  &   0.76  &   0.82  &   0.83    \\
& SDCT & 0.48 & 0.59 & 0.67 & 0.77 & 0.80 & 0.81 & 0.82 & 0.84 \\
Mean & BAS-2008 & 0.48 & 0.62 & 0.74 & 0.83 & 0.85 & 0.87 & 0.88 & 0.89 \\
SSIM & BAS-2009 & 0.48 & 0.62 & 0.73 & 0.82 & 0.84 & 0.85 & 0.87 & 0.88 \\
& BAS-2013 & 0.48 & 0.64 & 0.74 & 0.82 & 0.85 & 0.87 & 0.87 & 0.88 \\
& RDCT & 0.48 & 0.66 & 0.76 & 0.82 & 0.85 & 0.87 & 0.88 & 0.88 \\
& MRDCT & 0.48 & 0.55 & 0.65 & 0.72 & 0.76 & 0.83 & 0.84 & 0.86 \\
\midrule
& Exact DCT & 0.816  &  0.953  &  0.988  &  0.995  &  0.995   & 0.996  &  0.997  &  0.997  \\
& WHT &  0.815 &   0.815 &   0.823 &   0.823 &   0.955 &   0.955 &   0.982 &   0.982   \\
& SDCT &0.816  &  0.943   & 0.971   & 0.986   & 0.986  &  0.994  &  0.994   & 0.994  \\
Mean & BAS-2008 &0.816  &  0.936  &  0.973  &  0.993   & 0.993   & 0.993  &  0.993  &  0.995  \\
SR-SIM & BAS-2009&0.816  &  0.936  &  0.974  &  0.993   & 0.993   & 0.993  &  0.993  &  0.995  \\
& BAS-2013 & 0.815 &  0.951 &   0.982 &   0.997 &   0.997 &   0.997 &   0.997 &   0.997  \\
& RDCT &0.816 &   0.952 &   0.981 &   0.988 &   0.988 &   0.988 &   0.992 &   0.993  \\
& MRDCT &  0.816 &   0.898 &   0.932 &   0.958 &   0.958 &   0.982 &   0.986 &   0.988   \\
\bottomrule
\end{tabular}
\end{table*}

\subsection{Retained energy}

To further examine the performance of the pruned approximations,
we investigate
the signal energy distribution in
the transform-domain for each value of $K$.
This analysis is relevant,
because higher energy concentrations
implies that
$K$~can be reduced without severely degrading the
transform coding performance~\cite{britanak2007discrete}.
In fact,
higher energy concentration
effects
a large
number of zeros in the transform-domain
after quantization.
On its turn,
a large number of zeros
translates into longer runs of zeros,
which are beneficial for subsequent run-length encoding
and Huffman coding stages~\cite{bhaskaran1997}.

We analyzed a set of fifty $512 \times 512$
256-level grayscale
standard images from~\cite{USC_database}.
Originally color images were converted to grayscale by
extracting the luminance.
Image types included
textures,
satellite images,
landscapes,
portraits,
and natural images.
Such variety is to ensure that selection bias
is not introduced
in our experiments.
Thus our results are expected to be robust in this sense.
Images were split into 8$\times$8 subimages.
Resulting subimages
were submitted to each of the discussed
pruned DCT approximation for all values of~$K$.
Subsequently,
the relative amount of retained energy in the transform-domain
was computed.
Obtained values are displayed in Table~\ref{table-assessment}.

\subsection{Image Compression}

Proposed methods were submitted to an image compression simulation
to facilitate their performance as an image/video coding tool.
We based our
experiments
on
the
image compression
simulation
described in
in~\cite{Haweel2001,bas2008,Rao2001,bhaskaran1997,penn1992},
which is briefly outlined next.
We considered
the same above-mentioned set of
images,
sub-image decomposition,
and \mbox{2-D} pruned transformation,
as detailed in previous sub-section.
Resulting data were quantized
by dividing each term of the transformed matrix by elements of
the standard quantization matrix
for luminance~\cite[p.~153]{bhaskaran1997}.
Differently from~\cite{Haweel2001,bas2008,cb2011},
we included the quantization step in image compression simulation.
This is a more realistic and suitable approach
for pruned methods
which take advantage of quantization step.

\begin{figure*}%
\centering

 \begin{subfigure}[b]{0.24\linewidth}
        \includegraphics[scale=0.30]{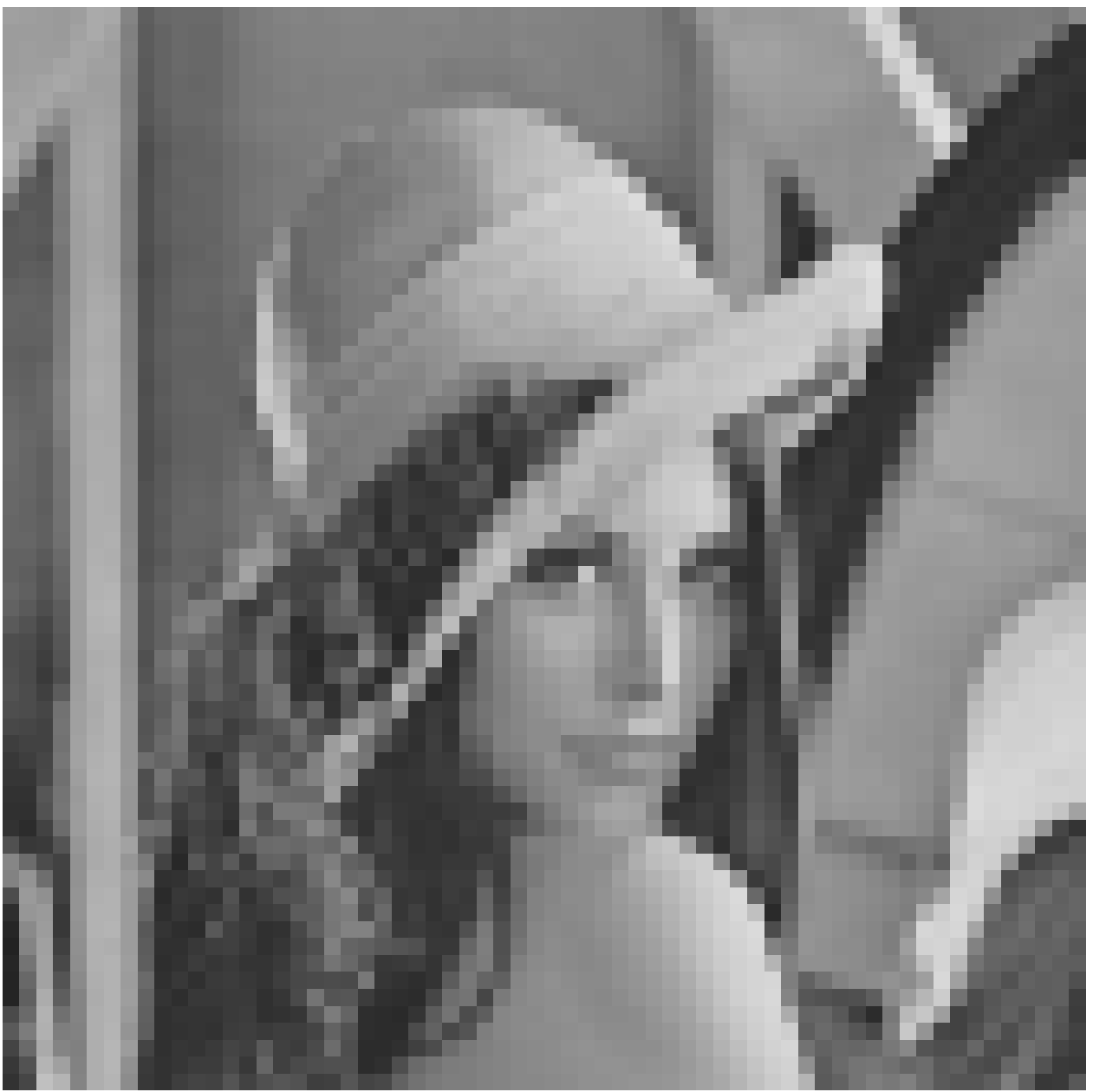}
        \caption{$K=1$ (PSNR=23.66, SSIM=0.63, \mbox{SR-SIM}=0.852)}
 \end{subfigure}
 \begin{subfigure}[b]{0.24\linewidth}
        \includegraphics[scale=0.30]{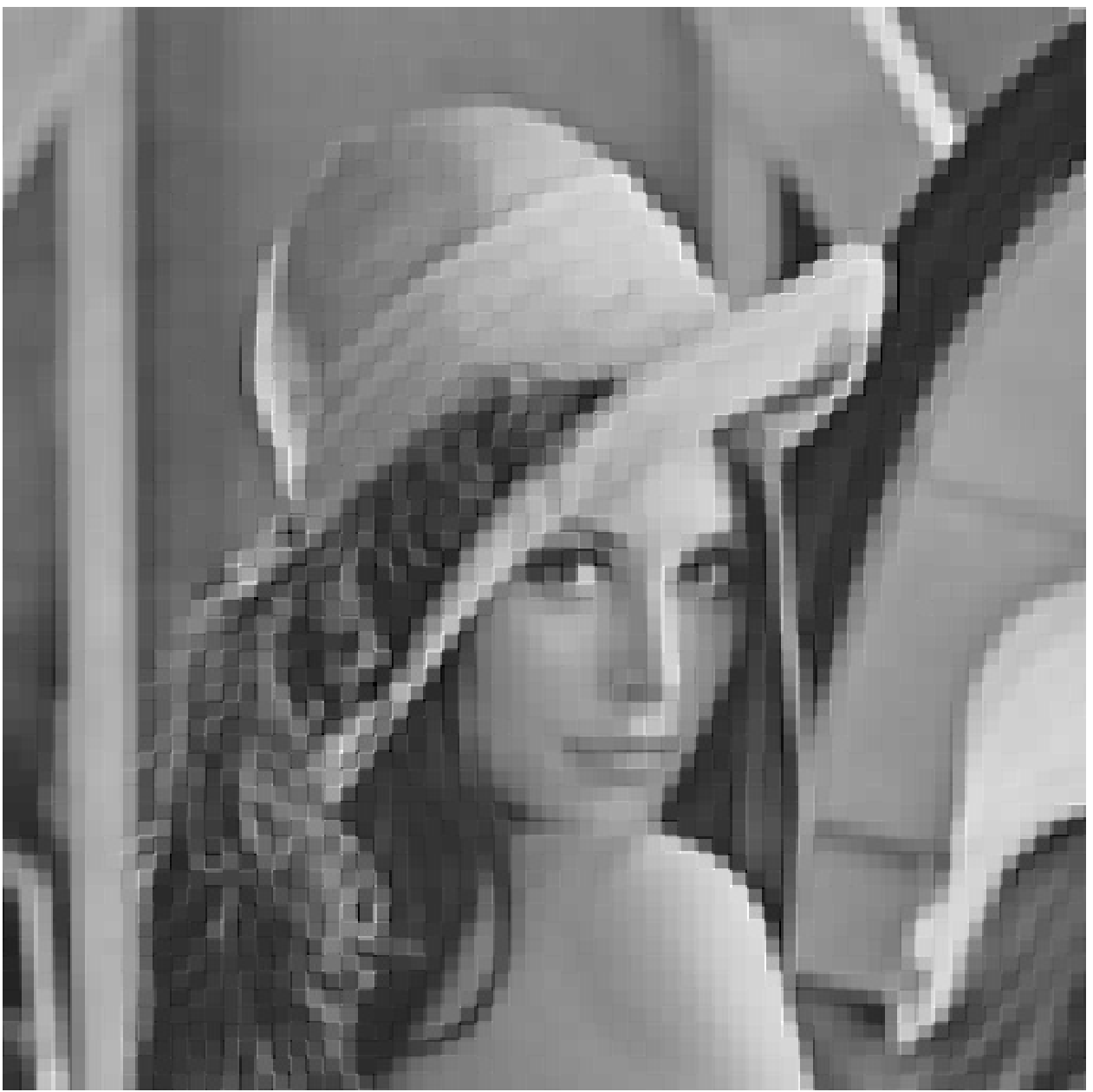}
        \caption{$K=2$ (PSNR=25.29, SSIM=0.69, \mbox{SR-SIM}=0.926)}
 \end{subfigure}
 \begin{subfigure}[b]{0.24\linewidth}
        \includegraphics[scale=0.30]{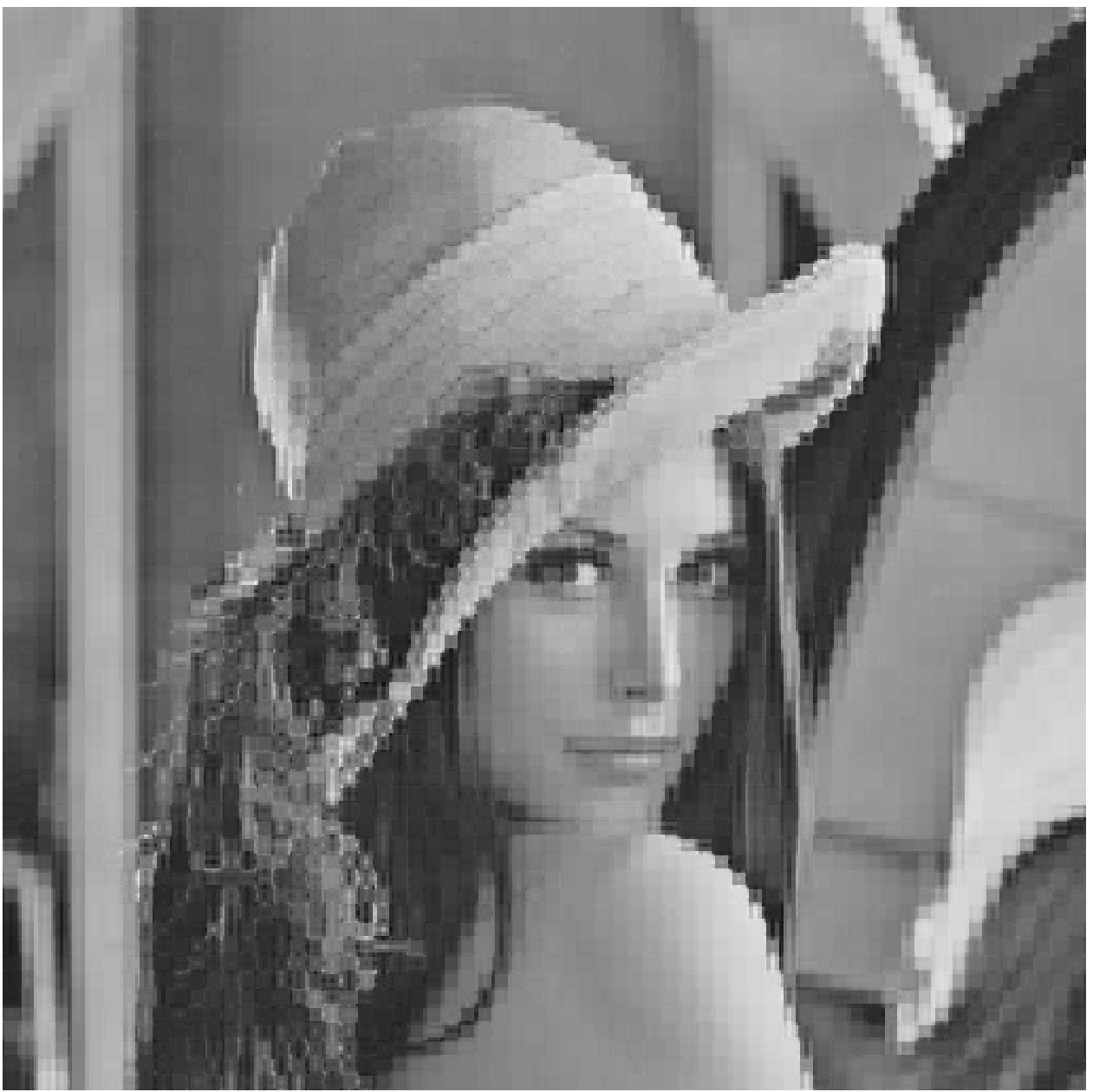}
        \caption{$K=3$ (PSNR=26.24, SSIM=0.75, \mbox{SR-SIM}=0.946)}
 \end{subfigure}
 \begin{subfigure}[b]{0.24\linewidth}
        \includegraphics[scale=0.30]{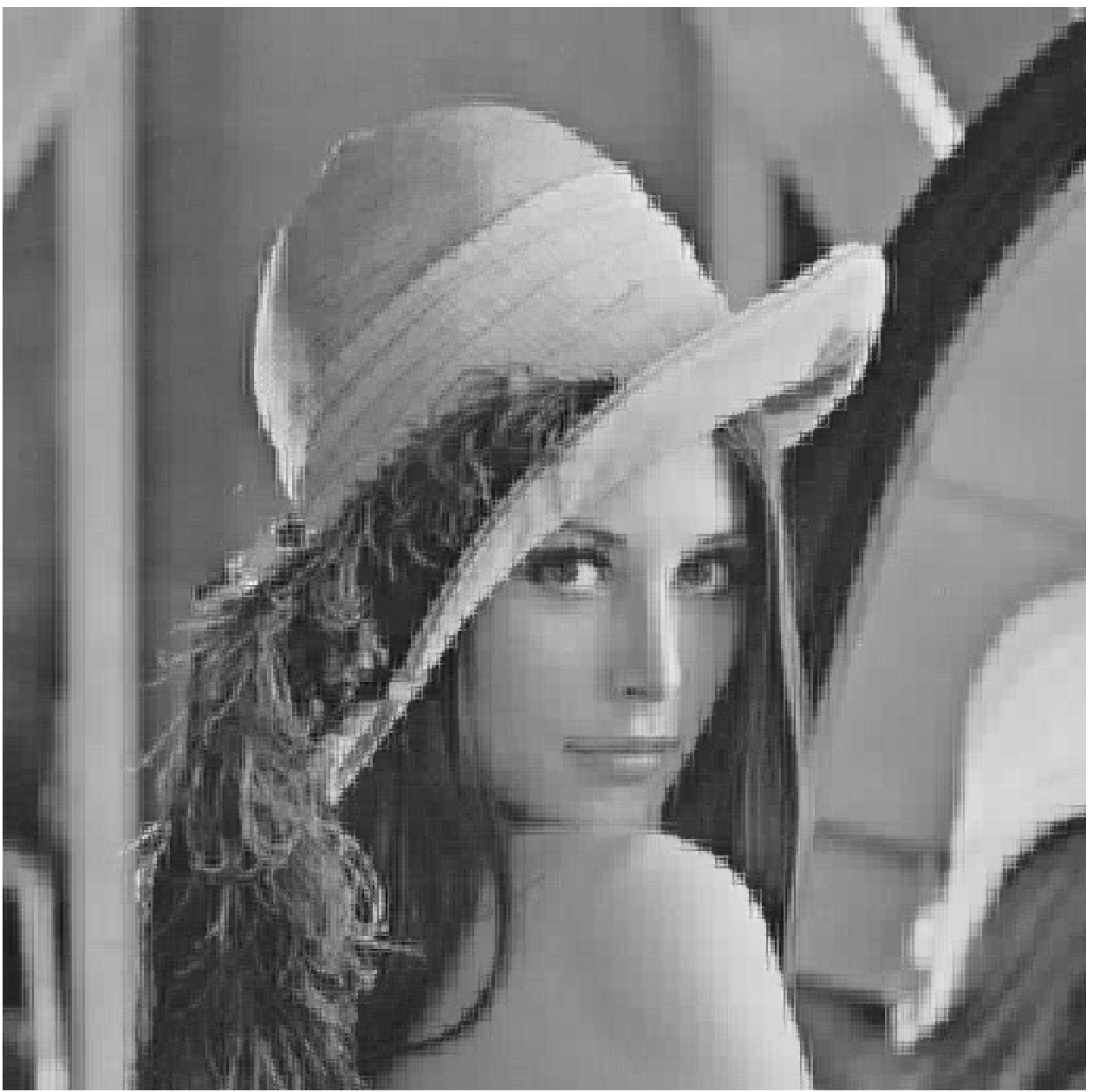}
        \caption{$K=4$ (PSNR=27.48, SSIM=0.79, \mbox{SR-SIM}=0.967)}
 \end{subfigure}
 \begin{subfigure}[b]{0.24\linewidth}
        \includegraphics[scale=0.30]{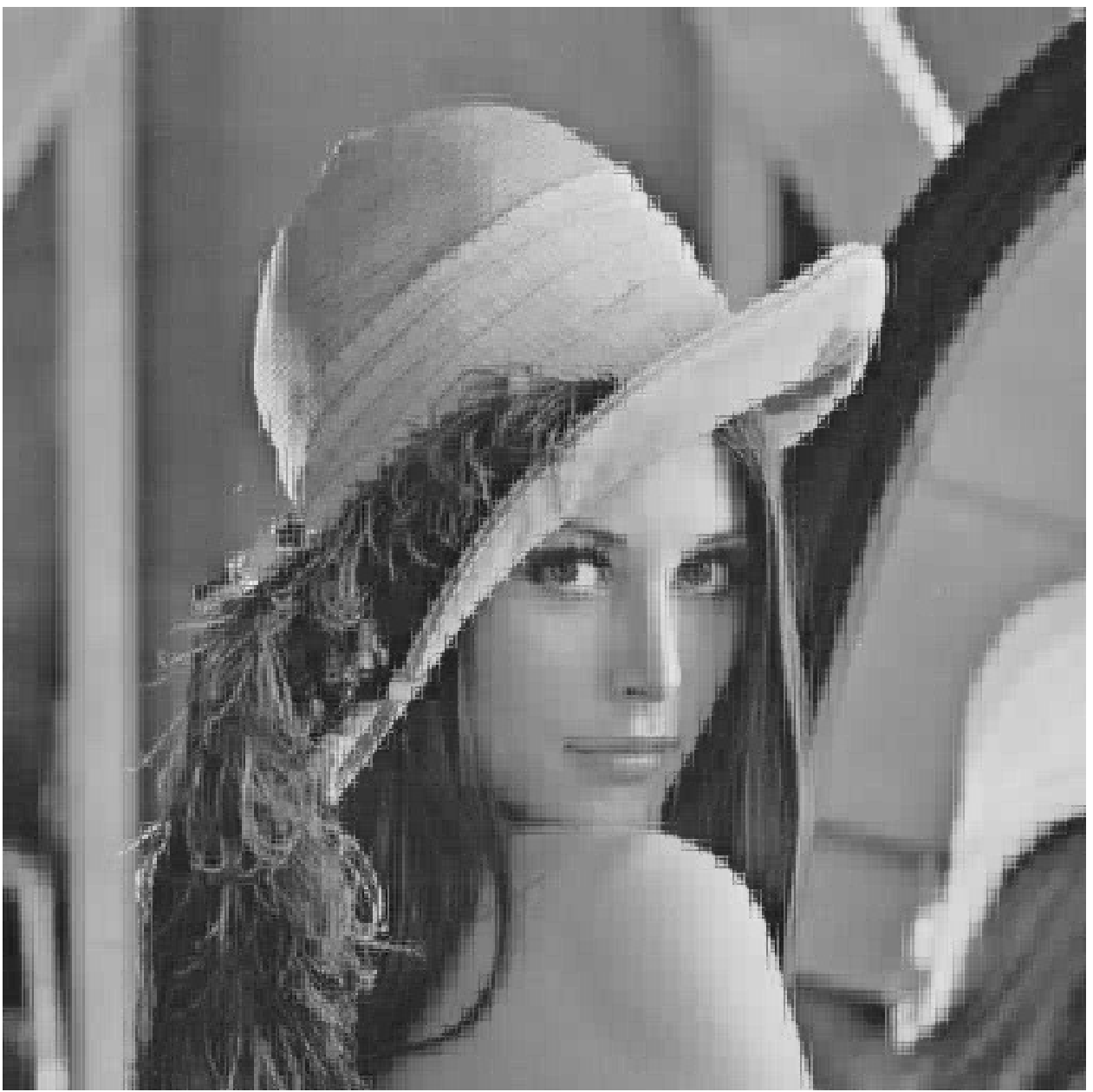}
        \caption{$K=5$ (PSNR=27.69, SSIM=0.80, \mbox{SR-SIM}=0.967)}
 \end{subfigure}
 \begin{subfigure}[b]{0.24\linewidth}
        \includegraphics[scale=0.30]{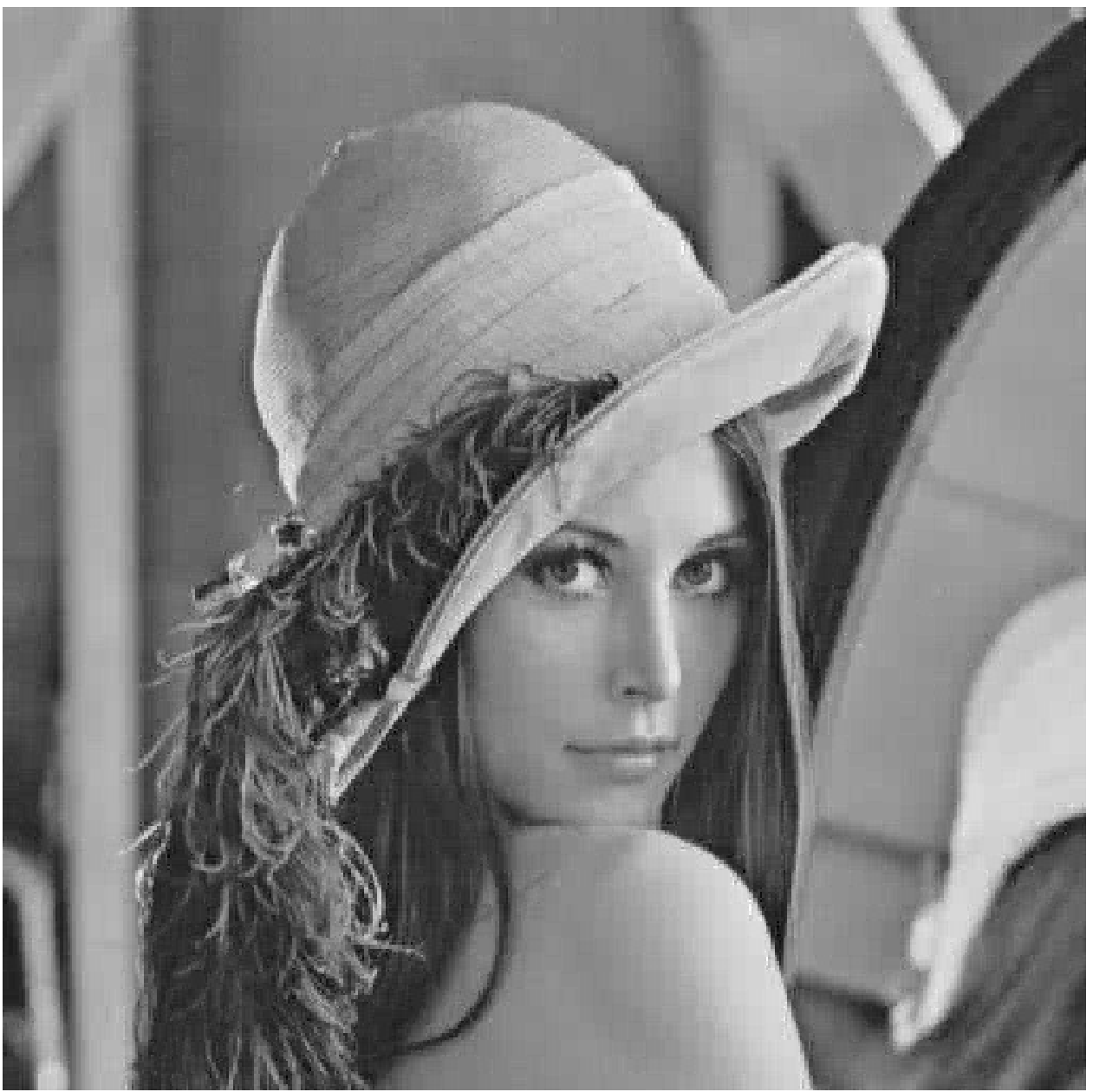}
        \caption{$K=6$ (PSNR=31.62, SSIM=0.86, \mbox{SR-SIM}=0.986)}
 \end{subfigure}
 \begin{subfigure}[b]{0.24\linewidth}
        \includegraphics[scale=0.30]{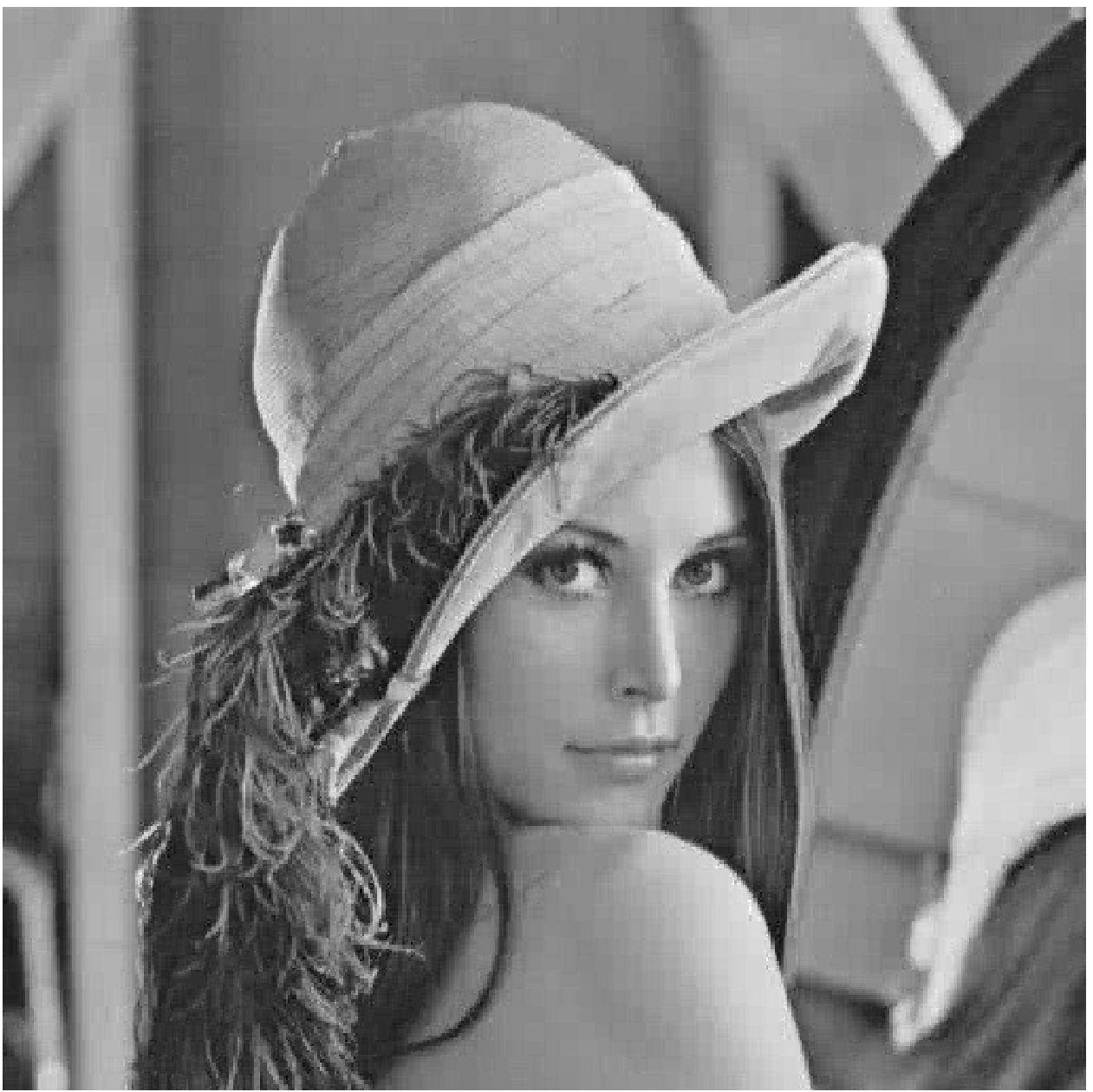}
        \caption{$K=7$ (PSNR=32.02, SSIM=0.87, \mbox{SR-SIM}=0.988)}
 \end{subfigure}
  \begin{subfigure}[b]{0.24\linewidth}
        \includegraphics[scale=0.30]{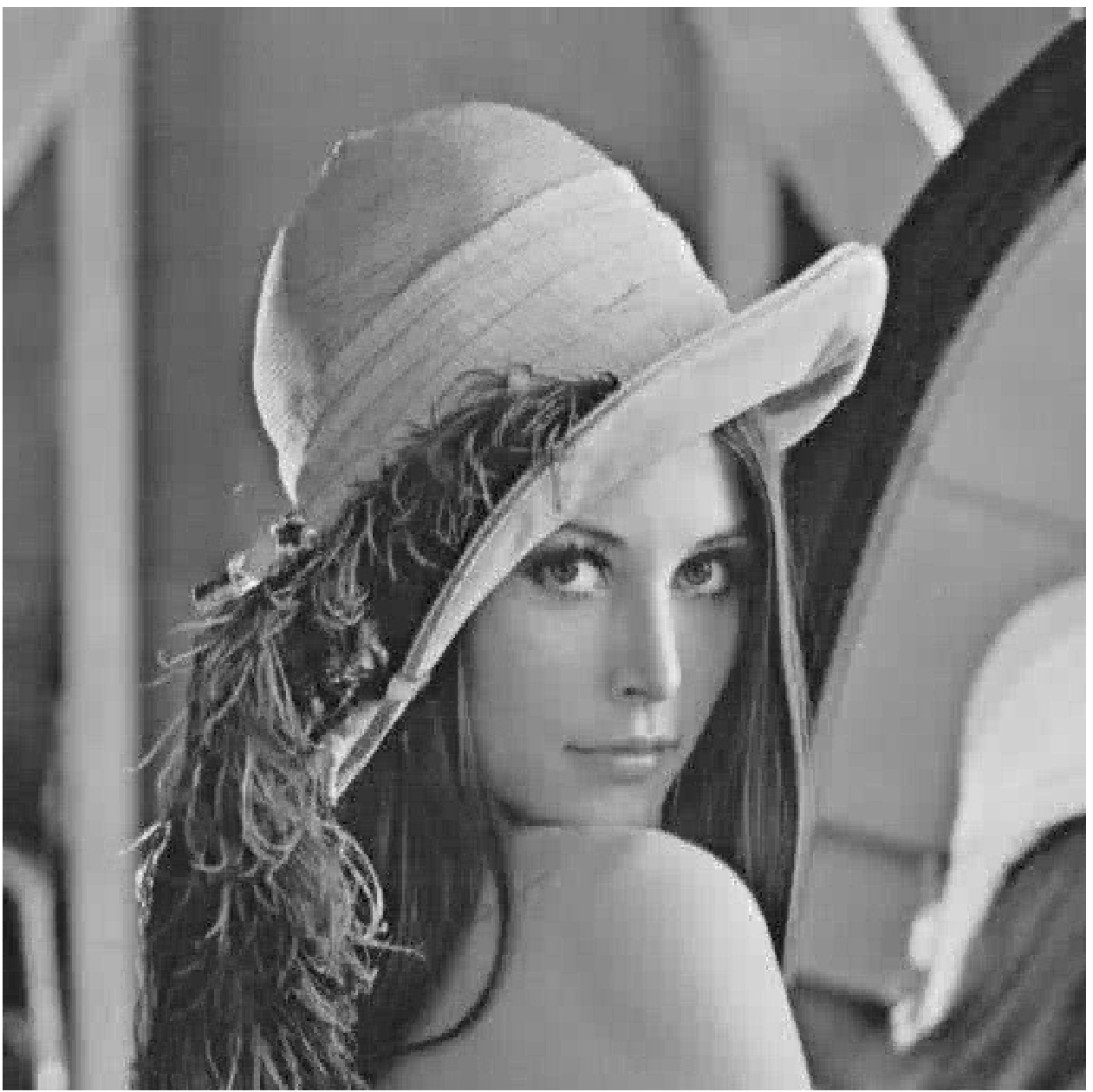}
        \caption{$K=8$ (PSNR=32.38, SSIM=0.87, \mbox{SR-SIM}=0.989)}
 \end{subfigure}
\caption{Reconstructed Lena image according to the pruned MRDCT}
\label{fig:lena}
\end{figure*}

An inverse procedure was applied
to reconstruct images considering
\mbox{2-D} inverse transform operation.
Recovered images were assessed for image degradation
by means
of peak signal-to-noise~(PSNR)~\cite[p.~9]{bhaskaran1997},
structural similarity index~(SSIM)~\cite{Wang2004},
and
spectral residual based similarity (SR-SIM)~\cite{sr-sim}.
The SSIM compares
an original image~$\mathbf{I}$
with
the recovered image~${\mathbf{R}}$
according to the following expression:
\begin{equation}
\operatorname{SSIM}\left(\mathbf{I}, \mathbf{R} \right) =
\frac{\left[ 2\mu_{_I}\mu_{_R} + \left( L\cdot10^{-2} \right) \right] \cdot \left[ 2\sigma_{_{IR}} +  \left(3L\cdot 10^{-2}\right) \right]}{\left[ \mu_{_I}^2 + \mu_{_R}^2 + \left(L\cdot 10^{-2}\right) \right]\cdot \left[\sigma_{_I}^2 + \sigma_{_R}^2 + \left(3L\cdot 10^{-2}\right) \right]},
\end{equation}
where
$
\mu_{_I} = \sum\limits_{i=1}^{8}\sum\limits_{j=1}^{8} \omega_{i,j} \cdot \mathbf{I}_{i,j}
$,
$
\sigma_{_I} = \sum\limits_{i=1}^{8}\sum\limits_{j=1}^{8} \omega_{i,j}\cdot \left( \mathbf{I}_{i,j} - \mu_{_I} \right)^{1/2}
$,
$
\sigma_{_{IR}} = \sum\limits_{i=1}^{8}\sum\limits_{j=1}^{8} \omega_{i,j}\cdot \left( \mathbf{I}_{i,j} - \mu_{_I} \right)\cdot \left( \mathbf{R}_{i,j} - \mu_{_R} \right)
$,
$L=255$ is the dynamic range of pixels
values,
and $\omega_{i,j}$ is entry of a Gaussian weighting function $\mathbf{w} = \left[\omega_{i,j}\right], i,j=1,2,\ldots,8$, with standard deviation of $1.5$ and normalized to unit sum.
The SR-SIM between the original image~$\mathbf{I}$ and the recovered image~${\mathbf{R}}$ is calculated as described in~\cite{sr-sim}.
Average PSNR, SSIM, and
SR-SIM
values
of all images were
computed
and are shown
in Table~\ref{table-assessment}.
For a qualitative analysis,
\figurename~\ref{fig:lena} displays
the reconstructed Lena image
computed via the MRDCT for all values of $K$.
Associated PSNR,
SSIM,
and
SR-SIM
values
are also shown.
Visual inspection suggests $K=6$
as good compromise between quality and
complexity.
Indeed,
we notice that the PSNR improvement from $K=5$ to $K=6$ is 3.92\,dB,
while the PSNR difference from $K=6$ and $K=7$ is just 0{.}4\,dB.

\subsection{Combined analysis}

In order to compare the discussed approximations,
we consider a combined figure of merit
that takes into account
some of the previously
discussed measures.
Although popular and worth reporting,
mean retained energy and PSNR are closely related measures.
Similarly,
the SR-SIM is a derivative of SSIM.
For a combined figure of merit,
we aim at selecting
unrelated measures;
thus
we separated
the
\mbox{2-D}
additive complexity,
PSNR,
and
SSIM
values,
whose numerical values
are listed in Table~\ref{table-assessment}.
Such combined measure is proposed
as
the following linear cost function:
\begin{equation}
\label{eq:cost}
\begin{split}
\operatorname{cost}(\mathbf{T}_K)
=
&
\alpha_1
\cdot
\operatorname{A}_\text{2-D}(\mathbf{T}_K)
+
(1-\alpha_1)
\cdot
\\
&
\Big\{
\alpha_2
\cdot
[
-\operatorname{NMSSIM}(\mathbf{T}_K)
]
+
(1 - \alpha_2)
\cdot
[
-\operatorname{NMPSNR}(\mathbf{T}_K)
]
\Big\}
,
\end{split}
\end{equation}
where
$\alpha_1,\alpha_2\in[0,1]$
are weights;
and
NMSSIM and NMPSNR
represent
the
normalized mean SSIM,
and
normalized mean PSNR,
respectively,
for all considered images
submitted to a particular
approximation~$\mathbf{T}_K$.
The above cost function consists of a multi-objective function,
which are commonly found in optimization literature~\cite{ehrgott2005multicriteria}.
Two types of metrics---arithmetic complexity
and
performance measurements---are subject to a convex combination
according to $\alpha_1$.
The performance measurements are themselves
a convex combination of SSIM and PSNR measurements,
balanced according to $\alpha_2$.
Thus the weights $\alpha_1$ and $\alpha_2$
control the relative importance of
the constituent metrics of the cost function.
For large values of $\alpha_1$,
the cost function emphasizes
the minimization of
the computational complexity;
whereas,
for
small values of $\alpha_1$,
the cost function
is proner to capture
measures of
image quality performance.
The quantity $\alpha_2$ balances
the composition of the performance measurement
between NMSSIM and NMPSNR.
Because
we consider $\alpha_1,\alpha_2\in[0,1]$,
all possible combinations of weights
are taken into account.
Only the particular context,
application,
and user requirements
can
determine the final choice of the weight values.

\begin{figure}
\centering
\begin{subfigure}{6cm}
\includegraphics[scale=1]{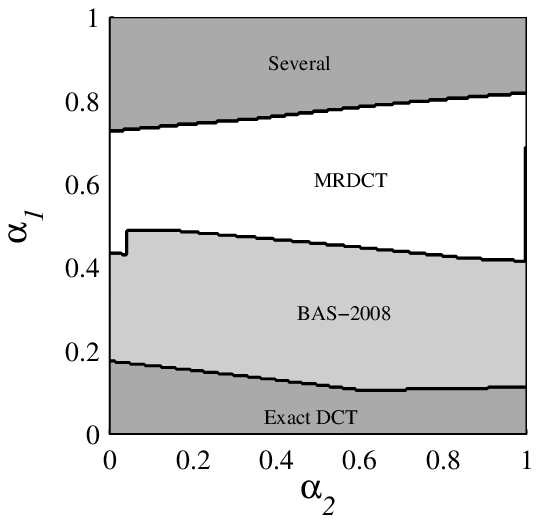}
\end{subfigure}
\begin{subfigure}{5cm}
\includegraphics[scale=1]{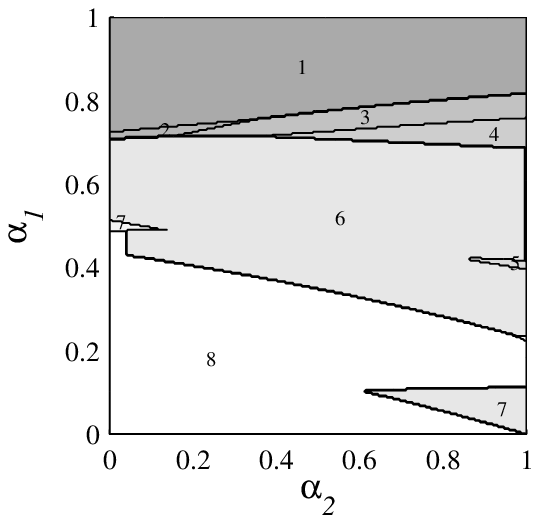}
\end{subfigure}

\caption{Optimality regions for the cost function:
(a)~optimal transform
and
(b)~optimal pruning value.}
\label{figure-cost}
\end{figure}

\figurename~\ref{figure-cost}(a) and~(b)
shows,
respectively,
regions
for
optimal transformation
and
pruning value $K$,
considering any choice of weights values $\alpha_1$ and $\alpha_2$.
For large $\alpha_1$ (emphasis in complex minimization),
the optimal choice tends to small $K$
regardless the transform.
Indeed,
for small $K$,
most pruned transformations collapse to the same
matrix.
For small $\alpha_1$ (emphasis in performance maximization),
optimality favors
more complex transformations
with large values of $K$,
being the full exact DCT the limiting case.

For mid-range values of $\alpha_1$,
we have less trival scenarios.
In \figurename~\ref{figure-cost}(a),
considering the optimal transform,
we notice that for mid-range values of $\alpha_1$
the MRDCT and the BAS-2008
occupy most of the
central area of the discussed region.
Around the same region
in
\figurename~\ref{figure-cost}(b),
for the MRDCT,
we obtain mostly $K=6$;
whereas
for the BAS-2008
we have $K=6,8$.
We emphasize that
the proposed pruned MRDCT with $K=6$
requires \emph{only 12 additions}.
The fast algorithm for this particular case
is presented in \figurename~\ref{figure-pruned-mrdct-k6}.

\subsection{HEVC Simulation}

\begin{figure}[h]
\centering
\begin{subfigure}{10cm}
\includegraphics[scale=1]{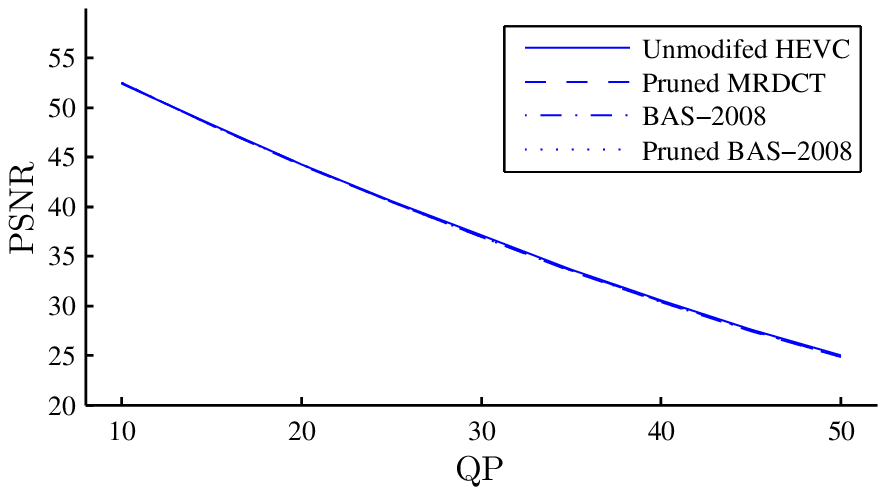}
\end{subfigure}
\\
\begin{subfigure}{10cm}
\includegraphics[scale=1]{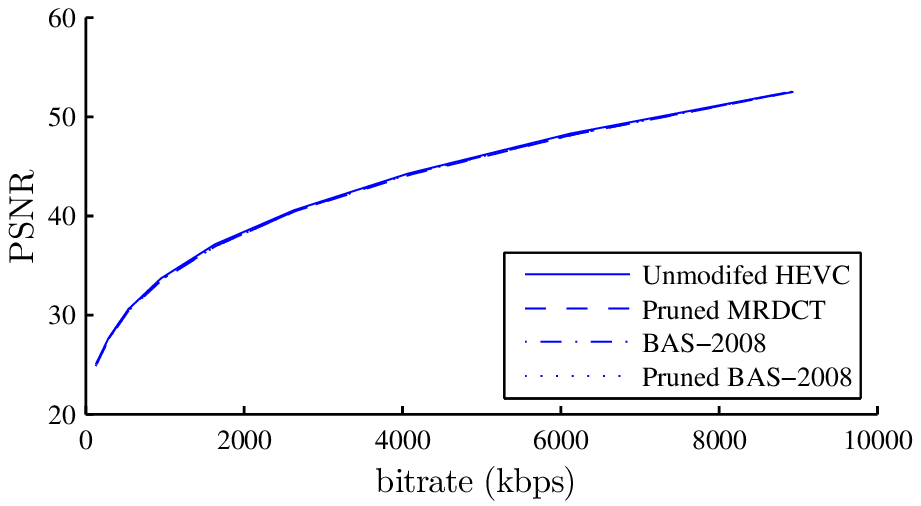}
\end{subfigure}

\caption{Video coding performance assessment.}
\label{figure-videocoding_quant}
\end{figure}

Taking into account the previous combined analysis,
we embedded
the proposed pruned MRDCT ($K=6$),
the BAS-2008 approximation ($K=8$),
and
the pruned BAS-2008 ($K=6$)
in the widely employed HEVC reference software HM~10.0~\cite{hm_software}.
This embedding consisted of substituting
the original 8-point integer-based DCT transform
present in the codec
for
each of the above-mentioned approximations.
We considered nine CIF video sequences
with 300~frames at 25~frames per second
from a public video bank~\cite{xiph_database}.
Such sequences were
submitted to encoding
according to:
(i)~the original software,
and
(ii)~the modified software.
We assessed mean PSNR metrics for luminance
by varying the quantization parameter~(QP)
from 10 to 50 with steps of 5~units.
Results are shown in
\figurename~\ref{figure-videocoding_quant}
considering both QP and bitrate.
Obtained curves are almost indistinguishable.
The mean PSNR values at $\text{QP}=30$
correspond to 37.06~dB, 36.92~dB, 36.96~dB, and 36.93~dB
for
the original integer DCT,
the pruned MRDCT ($K=6$),
BAS-2008,
and
the pruned BAS-2008 ($K=6$), respectively.
The degradation of the pruned approximations methods
relative to the unmodified software
was smaller than 0.15~dB for such QP value.

\begin{figure}[h]
\centering
\begin{subfigure}{10cm}
\includegraphics[scale=1]{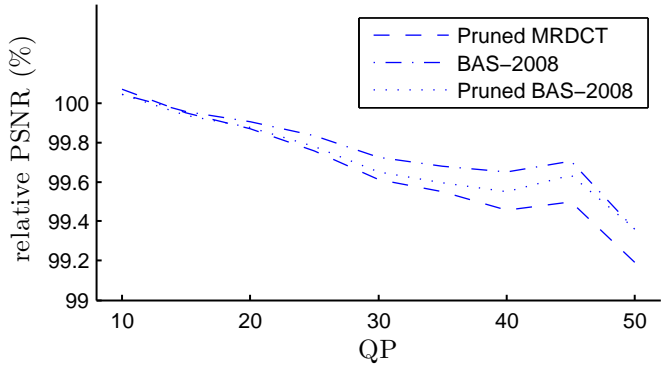}
\caption{Relative PSNR vs. QP}
\label{fig:relative_psnr_qp}
\end{subfigure}
\\
\begin{subfigure}{10cm}
\includegraphics[scale=1]{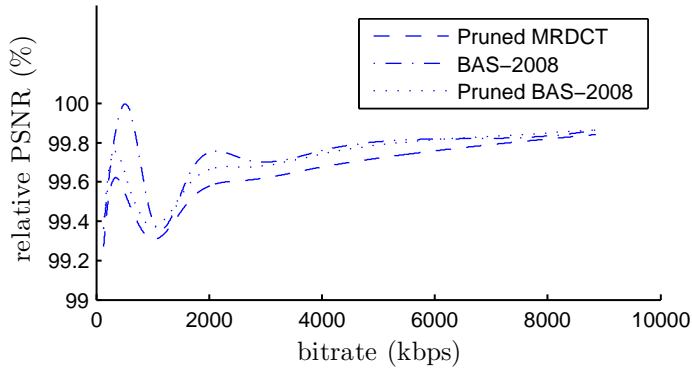}
\caption{Relative PSNR vs. Bitrate}
\label{fig:relative_psnr_br}
\end{subfigure}
\caption{Video coding performance assessment relative to Original HEVC.}
\label{fig:relative_psnr}
\end{figure}

\begin{figure*}

 \begin{subfigure}[b]{0.450\linewidth}
	\centering
        \includegraphics[scale=1.00]{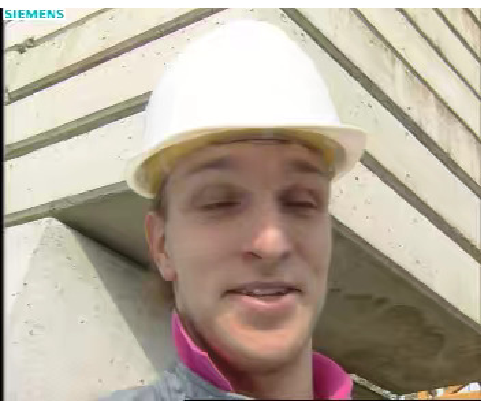}
        \caption{Unmodifed HEVC (PSNR=37.1154)}
 \end{subfigure}
 \begin{subfigure}[b]{0.450\linewidth}
	\centering
        \includegraphics[scale=1.00]{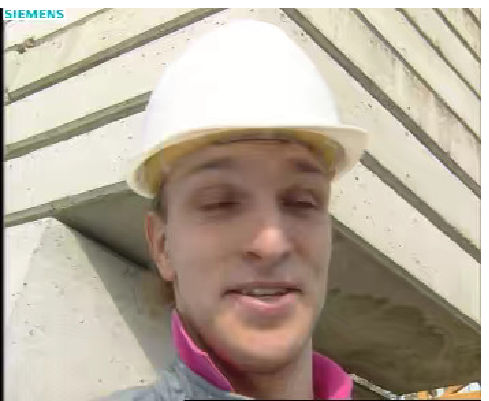}
        \caption{MRDCT $K=6$ (PSNR=37.0613)}
 \end{subfigure}
 \\
 \begin{subfigure}[b]{0.450\linewidth}
	\centering
        \includegraphics[scale=1.00]{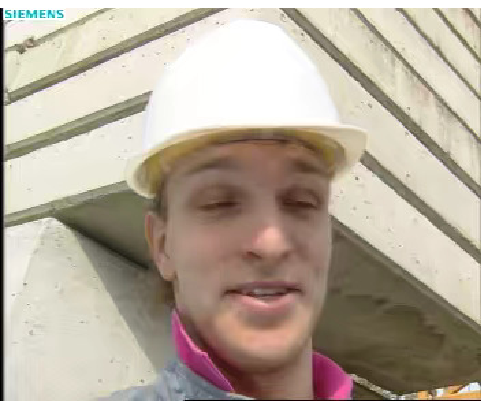}
        \caption{BAS-2008 $K=8$ (PSNR=37.0757)}
 \end{subfigure}
 \begin{subfigure}[b]{0.45\linewidth}
	\centering
        \includegraphics[scale=1.00]{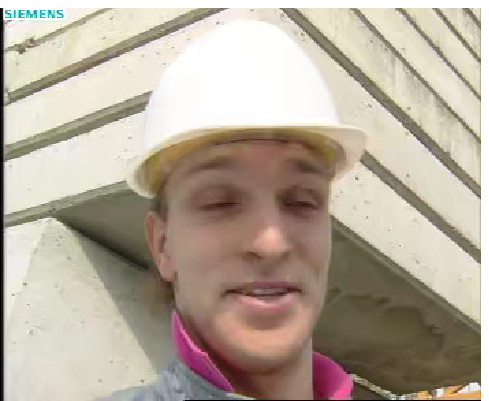}
        \caption{BAS-2008 $K=6$ (PSNR=37.0669)}
 \end{subfigure}
\caption{Reconstruct first frame of the ``Foreman'' video sequence encoded according to considered methods.}
\label{figure-videocoding_quali}
\end{figure*}

\figurename~\ref{fig:relative_psnr} shows
the relative percent PSNR of each approximate method
compared to the original HEVC according to QP and bitrate values.
The curves show very close performance to the original codec.
In \figurename~\ref{fig:relative_psnr_qp},
for low QP values,
the approximations show even higher PSNR, i.e.,
more than $100\%$ relative PSNR,
suggesting better compaction capability at low compression rates.
However,
same QP values do not necessarily
generate
the same compression ratio for each method,
since distinct coefficients are derived from each transformation and
submitted to
the same quantization table.
\figurename~\ref{fig:relative_psnr_br}
indicates
that the approximations
possess slightly
lower coding performance compared to original HEVC when compared at same bitrate.
At the same time,
the approximate methods present considerable lower computational cost
and
the lost of performance is smaller
than $1\%$.
\figurename~\ref{figure-videocoding_quali} shows
a qualitative comparison
considering
the first frame of the standard ``Foreman'' video sequence at $\mathrm{QP}=30$.
The degradation is hardly perceived.

\section{VLSI Architectures}
\label{sec:vlsi}

\begin{figure}
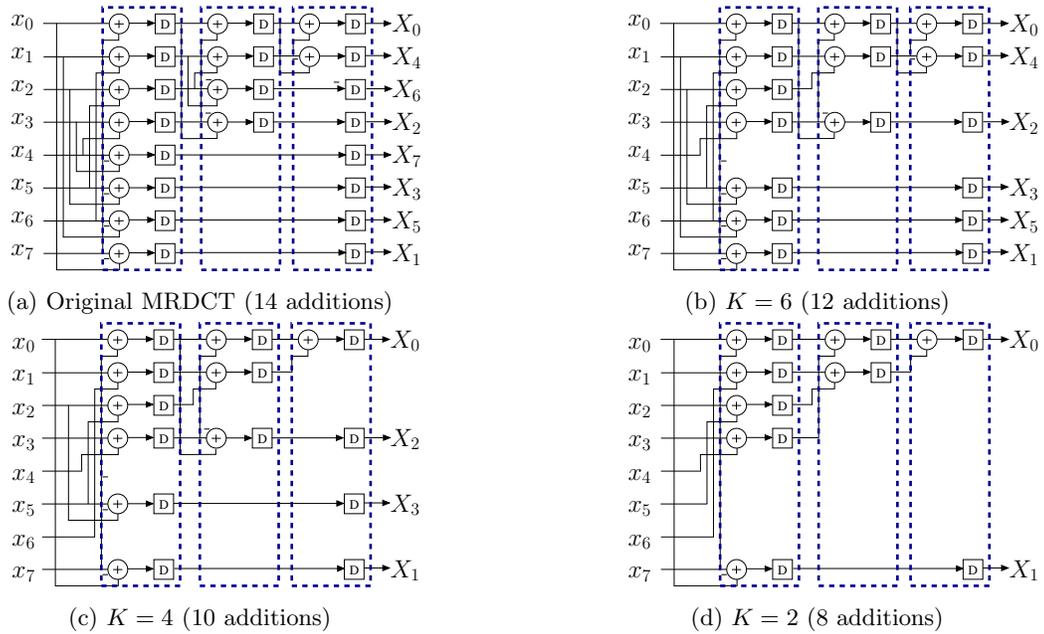
%
\centering
 \begin{subfigure}[b]{0.49\textwidth}
	\centering
        \scalebox{0.55}{\input{8point_n=8.tex}}
        \caption{Original MRDCT (14 additions)}
      \end{subfigure}
  \begin{subfigure}[b]{0.49\textwidth}
	\centering
         \scalebox{0.55}{\input{8point_n=6.tex}}
        \caption{$K = 6$ (12 additions)}
 \end{subfigure}
 \begin{subfigure}[b]{0.49\textwidth}
	\centering
         \scalebox{0.55}{\input{8point_n=4.tex}}
        \caption{$K = 4$ (10 additions)}
 \end{subfigure}
 \begin{subfigure}[b]{0.49\textwidth}
	\centering
         \scalebox{0.55}{\input{8point_n=2.tex}}
        \caption{$K = 2$ (8 additions)}
 \end{subfigure}

\caption{Digital architectures of the
MRDCT matrix and pruned MRDCT matrices for $K = 6,4,2$.}
\label{figure-mrdct-architectures}
\end{figure}

We aim at the physical realization
of
pruned designs
based on
the
MRDCT,
BAS-2008,
and
BAS-2013.
The MRDCT and BAS-2008 were selected
in accordance to the discussion in previous section.
The BAS-2013 was also included
because it is the base for
the only pruned approximate DCT competitor in literature~\cite{kouadria2013low}.
Such designs
were
realized in
a separable \mbox{2-D} block transform using
two \mbox{1-D} transform blocks with a transpose buffer between them.
Such blocks were designed and simulated,
using bit-true cycle-accurate modeling,
in Matlab/Simulink.
Thereafter,
the proposed architecture was ported to
Xilinx Virtex-6 field programmable gate array (FPGA)
as well as
to
custom CMOS standard-cell integrated circuit (IC) design.
The transform was applied in a row-parallel fashion
to the blocks of data
and
all blocks were $8 \times 8$,
irrespective of pruning.
When $K$ decreases, the number of
null elements in the blocks increases.
The row-transformed data were subject to transposition
and then
the same pruned algorithm was applied,
albeit for column direction.
\figurename~\ref{figure-mrdct-architectures}
shows the architectures for the MRDCT.
Remaining designs have similar realizations.

\subsection{FPGA Rapid Prototypes}

The pruned architectures
were physically realized on
a Xilinx Virtex-6 XC6VLX 240T-1FFG1156
FPGA
device with fine-grain pipelining for increased throughput.
The FPGA realizations were verified using
hardware-in-the-loop testing, which was achieved through a JTAG interface.
Proposed
approximations
were verified using more than 10000~test vectors
with complete agreement with theoretical values.
Evaluation of hardware complexity and real-time performance
considered the following metrics:
the number of employed configurable logic blocks (CLB),
flip-flop (FF) count,
critical path delay ($T_\text{cpd}$),
and
the maximum operating frequency ($F_{\text{max}}$) in~MHz.
The \texttt{xflow.results} report file,
from the Xilinx FPGA tool flow,
led to the reported results.
Frequency normalized dynamic power ($D_p$, in $\mathrm{mW}/\mathrm{MHz}$)
was
estimated using the Xilinx XPower Analyzer software tool.
Above measurements are shown in Table~\ref{fpga}
for the proposed pruned MRDCT
(highlighted in green),
the pruned version of the BAS-2008 introduced in~\cite{bas2008} (highlighted in blue)
and
the pruned BAS-2013 introduced in~\cite{bas2013}.

\begin{table}
\centering
\caption{Resource consumption on Xilinx XC6VLX240T-1FFG1156 device}
\label{fpga}
\begin{tabular}{lccccc}
\toprule
$K$ &
CLB &
FF &
$T_\text{cpd}$ &
$F_\text{max}$\! &
$D_p$\!
\\
\midrule
\multirow{3}{*}{1} & 107\cellcolor{green!50} & 376\cellcolor{green!50} & 2.263\cellcolor{green!50} & 441.89\cellcolor{green!50} & 0.67\cellcolor{green!50} \cellcolor{green!50} \\
& 107\cellcolor{blue!50} & 376\cellcolor{blue!50} & 2.263\cellcolor{blue!50} & 441.89\cellcolor{blue!50} & 0.67\cellcolor{blue!50}\cellcolor{blue!50} \\
& 107 & 376 & 2.263 & 441.89 & 0.67 \\
\hline
\multirow{3}{*}{2} & 136\cellcolor{green!50} & 568\cellcolor{green!50} & 2.300\cellcolor{green!50} & 434.78\cellcolor{green!50} & 0.97\cellcolor{green!50} \cellcolor{green!50} \\
& 203\cellcolor{blue!50} & 672\cellcolor{blue!50} & 2.600\cellcolor{blue!50} & 384.61\cellcolor{blue!50} & 1.33\cellcolor{blue!50} \cellcolor{blue!50} \\
& 204 & 751 & 2.450 & 408.10 & 1.45 \\
\hline
\multirow{3}{*}{3} & 210\cellcolor{green!50} & 783\cellcolor{green!50} & 2.509\cellcolor{green!50} & 398.56\cellcolor{green!50} & 0.87\cellcolor{green!50} \cellcolor{green!50} \\
& 252\cellcolor{blue!50} & 956\cellcolor{blue!50} & 2.878\cellcolor{blue!50} & 347.46\cellcolor{blue!50} & 1.74\cellcolor{blue!50} \cellcolor{blue!50} \\
& 263 & 978 & 2.534 & 394.63 & 2.11  \\
\hline
\multirow{3}{*}{4} & 247\cellcolor{green!50}  & 961\cellcolor{green!50} & 2.946\cellcolor{green!50} & 339.44\cellcolor{green!50} & 1.35\cellcolor{green!50} \cellcolor{green!50} \\
& 343\cellcolor{blue!50} & 1170\cellcolor{blue!50} & 3.100\cellcolor{blue!50} & 322.58\cellcolor{blue!50} & 2.06\cellcolor{blue!50} \cellcolor{blue!50} \\
& 339 & 1216 & 2.900 & 344.82 & 2.50 \\
\hline
\multirow{3}{*}{5} & 290\cellcolor{green!50} & 1123\cellcolor{green!50} & 2.877\cellcolor{green!50} & 347.58\cellcolor{green!50} & 1.70\cellcolor{green!50} \cellcolor{green!50} \\
& 362\cellcolor{blue!50} & 1331\cellcolor{blue!50} & 3.067\cellcolor{blue!50} & 326.05\cellcolor{blue!50} & 2.76\cellcolor{blue!50} \cellcolor{blue!50} \\
& 377 & 1374 & 2.902 & 344.58 & 3.13 \\
\hline
\multirow{3}{*}{6} & 350\cellcolor{green!50} & 1286\cellcolor{green!50} & 2.735\cellcolor{green!50} & 365.63\cellcolor{green!50} & 2.07\cellcolor{green!50} \cellcolor{green!50} \\
& 438\cellcolor{blue!50} & 1531\cellcolor{blue!50} & 3.214\cellcolor{blue!50} & 311.13\cellcolor{blue!50} & 3.07\cellcolor{blue!50} \cellcolor{blue!50} \\
& 382 & 1557 & 2.784 & 359.19 & 3.80  \\
\hline
\multirow{3}{*}{7} & 424\cellcolor{green!50} & 1487\cellcolor{green!50} & 3.300\cellcolor{green!50} & 303.03\cellcolor{green!50} & 2.21\cellcolor{green!50} \cellcolor{green!50} \\
& 501\cellcolor{blue!50} & 1709\cellcolor{blue!50} & 3.286\cellcolor{blue!50} & 304.32\cellcolor{blue!50} & 3.58\cellcolor{blue!50} \cellcolor{blue!50} \\
& 445 & 1720 & 3.432 & 291.37 & 3.87  \\
\hline
\multirow{3}{*}{8} & 445\cellcolor{green!50} & 1696\cellcolor{green!50} & 3.390\cellcolor{green!50} & 294.98\cellcolor{green!50} & 2.74\cellcolor{green!50} \cellcolor{green!50} \\
& 559\cellcolor{blue!50} & 1962\cellcolor{blue!50} & 3.300\cellcolor{blue!50} & 303.03\cellcolor{blue!50} & 3.85\cellcolor{blue!50} \cellcolor{blue!50} \\
& 517 & 1910 & 3.200 & 312.5 & 5.07  \\
\bottomrule
\end{tabular}

\end{table}

\subsection{ASIC Synthesis}

For the ASIC synthesis,
the hardware description
language code from the Xilinx System Generator FPGA design flow was ported to 45~nm CMOS technology and subject to
synthesis using Cadence Encounter.
Standard ASIC cells from the FreePDK, which
a free open-source cell library
at the 45~nm node,
was used for this purpose.
The supply voltage of the CMOS realization was fixed at
$V_\text{DD} = 1.1~\mathrm{V}$
during estimation of power consumption and logic delay.
The adopted figures of merit for the ASIC synthesis
were:
area ($A$) in~$\mathrm{mm^2}$,
area-time complexity ($AT$) in $\mathrm{mm}^2 \cdot \mathrm{ns}$,
area-time-squared complexity ($AT^2$) in $\mathrm{mm}^2 \cdot \mathrm{ns}^2$,
frequency normalized dynamic power ($D_p$, in $\mathrm{mW}/\mathrm{MHz}$),
critical path delay ($T_{cpd}$) in~$\mathrm{ns}$,
and
maximum operating frequency ($F_{\text{max}}$) in~GHz.
ASIC synthesis results
for the proposed pruned MRDCT (highlighted in green),
pruned version of the BAS-2008 (highlighted in blue)
and
the pruned BAS-2013 algorithm are displayed in Table~\ref{asic}.

\begin{table}
\centering
\caption{Resource consumption for CMOS 45\,nm ASIC synthesis}
\label{asic}
\begin{tabular}{lcccccc}
\toprule
$K$ &
Area &
AT &
${AT}^2$ &
$T_\text{cpd}$ &
$F_\text{max}$\! &
$D_p$\!
\\
\midrule
\multirow{3}{*}{1} & 0.011\cellcolor{green!50} & 0.011\cellcolor{green!50} & 0.010\cellcolor{green!50} & 0.961\cellcolor{green!50} & 1.040\cellcolor{green!50} & 0.018\cellcolor{green!50} \cellcolor{green!50} \\
& 0.011\cellcolor{blue!50} & 0.011\cellcolor{blue!50} & 0.010\cellcolor{blue!50} & 0.961\cellcolor{blue!50} & 1.040\cellcolor{blue!50} & 0.018\cellcolor{blue!50} \cellcolor{blue!50} \\
& 0.011 & 0.011 & 0.010 & 0.961 & 1.040 & 0.018 \\
\hline
\multirow{3}{*}{2} & 0.017\cellcolor{green!50} & 0.016\cellcolor{green!50} & 0.015\cellcolor{green!50} & 0.962\cellcolor{green!50} & 1.039\cellcolor{green!50} & 0.028\cellcolor{green!50} \cellcolor{green!50} \\
& 0.021\cellcolor{blue!50} & 0.020\cellcolor{blue!50} & 0.020\cellcolor{blue!50} & 0.980\cellcolor{blue!50} & 1.020\cellcolor{blue!50} & 0.035\cellcolor{blue!50} \cellcolor{blue!50} \\
& 0.022 & 0.022 & 0.022 & 0.995 & 1.005 & 0.036 \\
\hline
\multirow{3}{*}{3} & 0.022\cellcolor{green!50} & 0.021\cellcolor{green!50} & 0.020\cellcolor{green!50} & 0.963\cellcolor{green!50} & 1.038\cellcolor{green!50} & 0.038\cellcolor{green!50} \cellcolor{green!50} \\
& 0.031\cellcolor{blue!50} & 0.030\cellcolor{blue!50} & 0.030\cellcolor{blue!50} & 0.990\cellcolor{blue!50} & 1.010\cellcolor{blue!50} & 0.051\cellcolor{blue!50} \cellcolor{blue!50} \\
& 0.029 & 0.028 & 0.027 & 0.981 & 1.019 & 0.047 \\
\hline
\multirow{3}{*}{4} & 0.027\cellcolor{green!50} & 0.027\cellcolor{green!50} & 0.026\cellcolor{green!50} & 0.970\cellcolor{green!50} & 1.030\cellcolor{green!50} & 0.047\cellcolor{green!50} \cellcolor{green!50} \\
& 0.037\cellcolor{blue!50} & 0.037\cellcolor{blue!50} & 0.038\cellcolor{blue!50} & 1.016\cellcolor{blue!50} & 0.984\cellcolor{blue!50} & 0.063\cellcolor{blue!50} \cellcolor{blue!50} \\
& 0.037 & 0.036 & 0.036 & 0.997 & 1.003 & 0.059 \\
\hline
\multirow{3}{*}{5} & 0.032\cellcolor{green!50} & 0.034\cellcolor{green!50} & 0.037\cellcolor{green!50} & 1.075\cellcolor{green!50} & 0.930\cellcolor{green!50} & 0.057\cellcolor{green!50} \cellcolor{green!50} \\
& 0.042\cellcolor{blue!50} & 0.042\cellcolor{blue!50} & 0.043\cellcolor{blue!50} & 1.011\cellcolor{blue!50} & 0.989\cellcolor{blue!50} & 0.069\cellcolor{blue!50} \cellcolor{blue!50} \\
& 0.041 & 0.041 & 0.041 & 1.007 & 0.993 & 0.068  \\
\hline
\multirow{3}{*}{6} & 0.038\cellcolor{green!50} & 0.038\cellcolor{green!50} & 0.037\cellcolor{green!50} & 0.995\cellcolor{green!50} & 1.005\cellcolor{green!50} & 0.067\cellcolor{green!50} \cellcolor{green!50} \\
& 0.048\cellcolor{blue!50} & 0.048\cellcolor{blue!50} & 0.048\cellcolor{blue!50} & 1.000\cellcolor{blue!50} & 1.000\cellcolor{blue!50} & 0.081\cellcolor{blue!50} \cellcolor{blue!50} \\
& 0.046 & 0.046 & 0.046 & 1.008 & 0.992 & 0.077 \\
\hline
\multirow{3}{*}{7} & 0.043\cellcolor{green!50} & 0.047\cellcolor{green!50} & 0.051\cellcolor{green!50} & 1.085\cellcolor{green!50} & 0.921\cellcolor{green!50} & 0.079\cellcolor{green!50} \cellcolor{green!50} \\
& 0.053\cellcolor{blue!50} & 0.053\cellcolor{blue!50} & 0.054\cellcolor{blue!50} & 1.014\cellcolor{blue!50} & 0.986\cellcolor{blue!50} & 0.091\cellcolor{blue!50} \cellcolor{blue!50} \\
& 0.051 & 0.054 & 0.057 & 1.050 & 0.952 & 0.087 \\
\hline
\multirow{3}{*}{8} & 0.046\cellcolor{green!50} & 0.051\cellcolor{green!50} & 0.057\cellcolor{green!50} & 1.103\cellcolor{green!50} & 0.906\cellcolor{green!50} & 0.084\cellcolor{green!50} \cellcolor{green!50} \\
& 0.060\cellcolor{blue!50} & 0.062\cellcolor{blue!50} & 0.065\cellcolor{blue!50} & 1.047\cellcolor{blue!50} & 0.955\cellcolor{blue!50} & 0.104\cellcolor{blue!50} \cellcolor{blue!50} \\
& 0.057 & 0.057 & 0.058 & 1.008 & 0.992 & 0.097 \\
\bottomrule
\end{tabular}

\end{table}

\subsection{Discussion}%

The FPGA realization of the
proposed
pruned
MRDCT
showed
a drastic reductions in both area (measured from the number of CLBs) and frequency normalized dynamic power consumption, compared to the
full MRDCT.
Table~\ref{fpga-asic} shows
the percentage reduction of area and frequency-normalized
dynamic power for
both FPGA implementation and
CMOS synthesis for different pruning values.
All metrics indicate lower hardware resource consumption when the number of outputs are reduced from 8 to 1.
In particular, for $K=6$,
which minimizes the discussed cost function (cf.~\eqref{eq:cost}),
we notice a power consumption reduction
for approximately 20--25\%.

In order to compare the hardware resource consumption of
the introduced pruned DCT approximation with competing transforms,
we physically realized the pruned BAS-2013 algorithm~\cite{bas2013} and the pruned BAS-2008 algorithm~\cite{bas2008}
on the same Xilinx Virtex-6 XC6VLX240T-1FFG1156 device
and
submitted it to synthesis using ASIC 45~nm CMOS technology.
By comparing the results in Table~\ref{fpga} and~\ref{asic},
it can be seen that the proposed transform
discussed here
outperforms both pruned BAS-2008 and pruned BAS-2013
in terms of hardware resource consumption,
and power consumption while is in par in terms of speed as well.

\begin{table}
\centering
\caption{Percentage reduction in area and dynamic power for FPGA}
\label{fpga-asic}
\begin{tabular}{lcc | cc}
\toprule
 &
\multicolumn{2}{ c| }{FPGA} &
\multicolumn{2}{ c }{ASIC} \\
\midrule
$K$ &
Area \% &
$D_{p}$ \% &
Area \% &
$D_{p}$ \%
\\
\midrule
1 & 71.65 & 83.11 & 75.32 & 76.66 \\
2 & 54.59 & 72.29 & 64.93 & 66.00 \\
3 & 44.88 & 62.33 & 53.24 & 54.66 \\
4 & 30.18 & 51.94 & 41.55 & 43.33  \\
5 & 19.16 & 34.63 & 32.46 & 34.00 \\
6 & 3.14 & 20.77 & 23.37 & 24.66 \\
7 & 1.57 & 12.12 & 10.38 & 12.66 \\
\bottomrule
\end{tabular}

\end{table}

\section{Conclusion}
\label{sec:conclusion}

In this paper,
we present a set of
8-point pruned DCT approximations
derived
from state-of-the-art methods.
All possible frequency-domain pruning schemes
were considered and analyzed
in terms of arithmetic complexity,
energy compaction in the transform-domain,
and image compression performance.
A new combined metric was defined considering
the \mbox{2-D} arithmetic complexity and
average values of PSNR and SSIM.
The pruned transform based on MRDCT presented
the lowest arithmetic complexity and
the showed competitive performance.
Thus,
the pruned MRDCT approximations were digitally implemented
using both Xilinx FPGA tools and CMOS 45 nm
ASIC technology.
The proposed pruned transforms
demonstrated practical relevance
in image/video compression.
The proposed algorithms are
fully compatible with modern codecs.
We have embedded
the proposed methods
into a standard HEVC reference software~\cite{hm_software}.
Results presented very low qualitative and quantitative degradation
at a considerable lower computational cost.

Additionally,
low-complexity designs
are required in
several contexts
were very high quality imagery is not
a strong requirement,
such as:
environmental monitoring, habitat
monitoring, surveillance, structural monitoring,
equipment diagnostics, disaster management, and
emergency response~\cite{Kimura2005}.
All above contexts can benefit of
the proposed tools
when embedded into wireless sensors
with low-complexity codecs and low-power hardware~\cite{WMSN2007survey}.

We summarize the contributions of the present work:
\begin{itemize}
\item
The pruning approach for DCT approximations
was generalized
by not only considering all possible pruning variations
but also
investigating a wide range of DCT approximations;

\item
An analysis covering all cases
under different
figures of merit,
including
arithmetic complexity
and
image quality measures
was presented;

\item
A combined figure of merit
to guide the decision making process
in terms hardware realization was introduced;

\item
The \mbox{2-D} case
was also analyzed
and
concluded that the pruning approach is
even better suited for \mbox{2-D} transforms.

\item
The considered pruned DCT approximation
was implemented
using Xilinx FPGA tools and
synthesized using CMOS 45~nm ASIC technology.
Such implementations demonstrated the low resource consumption
of the proposed pruned transform.
\end{itemize}

\section*{Acknowledgements}

This work was partially supported by
CNPq, FACEPE, and FAPERGS (Brazil),
and by the College of Engineering at the University of Akron, Akron, OH, USA.

{\small
\bibliographystyle{IEEEtran}
\bibliography{dct}

\begin{thebibliography}{10}
\providecommand{\url}[1]{#1}
\csname url@samestyle\endcsname
\providecommand{\newblock}{\relax}
\providecommand{\bibinfo}[2]{#2}
\providecommand{\BIBentrySTDinterwordspacing}{\spaceskip=0pt\relax}
\providecommand{\BIBentryALTinterwordstretchfactor}{4}
\providecommand{\BIBentryALTinterwordspacing}{\spaceskip=\fontdimen2\font plus
\BIBentryALTinterwordstretchfactor\fontdimen3\font minus
  \fontdimen4\font\relax}
\providecommand{\BIBforeignlanguage}[2]{{%
\expandafter\ifx\csname l@#1\endcsname\relax
\typeout{** WARNING: IEEEtran.bst: No hyphenation pattern has been}%
\typeout{** loaded for the language `#1'. Using the pattern for}%
\typeout{** the default language instead.}%
\else
\language=\csname l@#1\endcsname
\fi
#2}}
\providecommand{\BIBdecl}{\relax}
\BIBdecl

\bibitem{ahmed1975}
N.~Ahmed and K.~R. Rao, \emph{Orthogonal Transforms for Digital Signal
  Processing}.\hskip 1em plus 0.5em minus 0.4em\relax Springer, 1975.

\bibitem{Blahut2010}
R.~E. Blahut, \emph{Fast Algorithms for Signal Processing}.\hskip 1em plus
  0.5em minus 0.4em\relax Cambridge University Press, 2010.

\bibitem{Wallace1992}
G.~Wallace, ``The {JPEG} still picture compression standard,'' \emph{IEEE
  Transactions on Consumer Electronics}, vol.~38, no.~1, pp. xviii--xxxiv,
  1992.

\bibitem{Gall1992}
D.~J.~L. Gall, ``The {MPEG} video compression algorithm,'' \emph{Signal
  Processing: Image Communication}, vol.~4, pp. 129--140, 1992.

\bibitem{roma2007hybrid}
N.~Roma and L.~Sousa, ``Efficient hybrid {DCT}-domain algorithm for video
  spatial downscaling,'' \emph{EURASIP Journal on Advances in Signal
  Processing}, vol. 2007, no.~2, pp. 30--30, 2007.

\bibitem{mpeg2}
{International Organisation for Standardisation}, ``Generic coding of moving
  pictures and associated audio information -- {P}art 2: Video,'' ISO, {ISO/IEC
  JTC1/SC29/WG11} - Coding of Moving Pictures and Audio, 1994.

\bibitem{h261}
{International Telecommunication Union}, ``{ITU}-{T} recommendation {H}.261
  version 1: Video codec for audiovisual services at $p \times 64$ kbits,''
  ITU-T, Tech. Rep., 1990.

\bibitem{Liou1990}
M.~L. Liou, ``Visual telephony as an {ISDN} application,'' \emph{{IEEE}
  Communications Magazine}, vol.~28, pp. 30--38, 1990.

\bibitem{h263}
{International Telecommunication Union}, ``{ITU}-{T} recommendation {H}.263
  version 1: Video coding for low bit rate communication,'' ITU-T, Tech. Rep.,
  1995.

\bibitem{wiegand2003}
T.~Wiegand, G.~J. Sullivan, G.~Bjontegaard, and A.~Luthra, ``Overview of the
  {H}.264/{AVC} video coding standard,'' \emph{IEEE Transactions on Circuits
  and Systems for Video Technology}, vol.~13, no.~7, pp. 560--576, Jul. 2003.

\bibitem{h264}
J.~V. Team, ``Recommendation {H}.264 and {ISO}/{IEC} 14 496--10 {AVC}: Draft
  {ITU}-{T} recommendation and final draft international standard of joint
  video specification,'' ITU-T, Tech. Rep., 2003.

\bibitem{hevc}
{International Telecommunication Union}, ``High efficiency video coding:
  Recommendation {ITU-T H.265},'' ITU-T Series H: Audiovisual and Multimedia
  Systems, Tech. Rep., 2013.

\bibitem{hevc1}
M.~T. Pourazad, C.~Doutre, M.~Azimi, and P.~Nasiopoulos, ``{HEVC}: The new gold
  standard for video compression: How does {HEVC} compare with {H.264/AVC}?''
  \emph{IEEE Consumer Electronics Magazine}, vol.~1, no.~3, pp. 36--46, Jul.
  2012.

\bibitem{Park2012}
J.-S. Park, W.-J. Nam, S.-M. Han, and S.~Lee, ``2-{D} large inverse transform
  (16$\times$16, 32$\times$32) for {HEVC} ({H}igh {E}fficiency {V}ideo
  {C}oding),'' \emph{Journal of Semiconductor Technology and Science}, vol.~2,
  pp. 203--211, 2012.

\bibitem{Ohm2012}
J.-R. Ohm, G.~J. Sullivan, H.~Schwarz, T.~K. Tan, and T.~Wiegand, ``Comparison
  of the coding efficiency of video coding standards - including {H}igh
  {E}fficiency {V}ideo {C}oding ({HEVC}),'' \emph{IEEE Transactions on Circuits
  and Systems for Video Technology}, vol.~22, no.~12, pp. 1669--1684, Dec.
  2012.

\bibitem{Potluri2013}
U.~S. Potluri, A.~Madanayake, R.~J. Cintra, F.~M. Bayer, S.~Kulasekera, and
  A.~Edirisuriya, ``Improved 8-point approximate {DCT} for image and video
  compression requiring only 14 additions,'' \emph{IEEE Transactions on
  Circuits and Systems I}, vol.~61, no.~6, pp. 1727--1740, 2014.

\bibitem{sullivan2012}
G.~J. Sullivan, J.-R. Ohm, W.-J. Han, and T.~Wiegand, ``Overview of the high
  efficiency video coding ({HEVC}) standard,'' \emph{IEEE Transactions on
  Circuits and Systems for Video Technology}, vol.~22, no.~12, pp. 1649--1668,
  Dec. 2012.

\bibitem{Chen1977}
W.~H. Chen, C.~Smith, and S.~Fralick, ``A fast computational algorithm for the
  discrete cosine transform,'' \emph{IEEE Transactions on Communications},
  vol.~25, no.~9, pp. 1004--1009, Sep. 1977.

\bibitem{hou1987fast}
H.~S. Hou, ``A fast recursive algorithm for computing the discrete cosine
  transform,'' \emph{IEEE Transactions on Acoustic, Signal, and Speech
  Processing}, vol.~6, no.~10, pp. 1455--1461, 1987.

\bibitem{Arai1988}
Y.~Arai, T.~Agui, and M.~Nakajima, ``A fast {DCT}-{SQ} scheme for images,''
  \emph{Transactions of the IEICE}, vol. E-71, no.~11, pp. 1095--1097, Nov.
  1988.

\bibitem{Loeffler1989}
C.~Loeffler, A.~Ligtenberg, and G.~S. Moschytz, ``Practical fast 1-{D DCT}
  algorithms with 11 multiplications,'' \emph{{ICASSP} International Conference
  on Acoustics, Speech, and Signal Processing}, vol.~2, pp. 988--991, 1989.

\bibitem{fw1992}
E.~Feig and S.~Winograd, ``Fast algorithms for the discrete cosine transform,''
  \emph{IEEE Transactions on Signal Processing}, vol.~40, no.~9, pp.
  2174--2193, 1992.

\bibitem{britanak2007discrete}
V.~Britanak, P.~Yip, and K.~R. Rao, \emph{Discrete Cosine and Sine
  Transforms}.\hskip 1em plus 0.5em minus 0.4em\relax Academic Press, 2007.

\bibitem{winograd1980}
S.~Winograd, \emph{Arithmetic Complexity of Computations}.\hskip 1em plus 0.5em
  minus 0.4em\relax CBMS-NSF Regional Conference Series in Applied Mathematics,
  1980.

\bibitem{Haweel2001}
T.~I. Haweel, ``A new square wave transform based on the {DCT},'' \emph{Signal
  Processing}, vol.~82, pp. 2309--2319, 2001.

\bibitem{lengwehasatit2004scalable}
K.~Lengwehasatit and A.~Ortega, ``Scalable variable complexity approximate
  forward {DCT},'' \emph{IEEE Transactions on Circuits and Systems for Video
  Technology}, vol.~14, no.~11, pp. 1236--1248, Nov. 2004.

\bibitem{bas2008}
S.~Bouguezel, M.~O. Ahmad, and M.~N.~S. Swamy, ``Low-complexity 8$\times$8
  transform for image compression,'' \emph{Electronics Letters}, vol.~44,
  no.~21, pp. 1249--1250, Sep. 2008.

\bibitem{cb2011}
R.~J. Cintra and F.~M. Bayer, ``A {DCT} approximation for image compression,''
  \emph{IEEE Signal Processing Letters}, vol.~18, no.~10, pp. 579--582, Oct.
  2011.

\bibitem{bas2009}
S.~Bouguezel, M.~O. Ahmad, and M.~N.~S. Swamy, ``A fast 8$\times$8 transform
  for image compression,'' in \emph{2009 International Conference on
  Microelectronics (ICM)}, Dec. 2009, pp. 74--77.

\bibitem{bas2013}
------, ``Binary discrete cosine and {H}artley transforms,'' \emph{IEEE
  Transactions on Circuits and Systems I: Regular Papers}, vol.~60, no.~4, pp.
  989--1002, 2013.

\bibitem{Cintra2014-sigpro}
R.~J. Cintra, F.~M. Bayer, and C.~J. Tablada, ``Low-complexity 8-point {DCT}
  approximations based on integer functions,'' \emph{Signal Processing},
  vol.~99, pp. 201--214, 2014.

\bibitem{Makkaoui2010}
L.~Makkaoui, V.~Lecuire, and J.~Moureaux, ``Fast zonal {DCT}-based image
  compression for wireless camera sensor networks,'' \emph{2nd International
  Conference on Image Processing Theory Tools and Applications ({IPTA})}, pp.
  126--129, 2010.

\bibitem{Docef2002}
A.~Docef, ``The quantized {DCT} and its application to {DCT}-based video
  coding,'' \emph{{IEEE} Transactions on Image Processing}, vol.~11, pp.
  177--187, 2002.

\bibitem{rao1990discrete}
K.~R. Rao and P.~Yip, \emph{Discrete Cosine Transform: Algorithms, Advantages,
  Applications}.\hskip 1em plus 0.5em minus 0.4em\relax San Diego, CA: Academic
  Press, 1990.

\bibitem{Ahmed1974}
N.~Ahmed, T.~Natarajan, and K.~R. Rao, ``Discrete cosine transform,''
  \emph{IEEE Transactions on Computers}, vol. C-23, no.~1, pp. 90--93, Jan.
  1974.

\bibitem{Rao2001}
K.~R. Rao and P.~Yip, \emph{The Transform and Data Compression Handbook}.\hskip
  1em plus 0.5em minus 0.4em\relax {CRC} Press {LLC}, 2001.

\bibitem{Malepati2010}
H.~Malepati, \emph{Digital Media Processing: {DSP} Algorithms Using {C} (Google
  e-Livro)}.\hskip 1em plus 0.5em minus 0.4em\relax Newnes, 2010.

\bibitem{Huang2000}
Y.-M. Huang, J.-L. Wu, and C.-L. Chang, ``A generalized output pruning
  algorithm for matrix-vector multiplication and its application to compute
  pruning discrete cosine transform,'' \emph{IEEE Transactions on Signal
  Processing}, vol.~48, pp. 561--563, 2000.

\bibitem{wang2012generic}
L.~Wang, X.~Zhou, G.~Sobelman, and R.~Liu, ``Generic mixed-radix {FFT}
  pruning,'' \emph{IEEE Signal Processing Letters}, vol.~19, no.~3, pp.
  167--170, March 2012.

\bibitem{airoldi2010energy}
R.~Airoldi, O.~Anjum, F.~Garzia, A.~M. Wyglinski, and J.~Nurmi,
  ``Energy-efficient fast {F}ourier transforms for cognitive radio systems,''
  \emph{IEEE Micro}, vol.~30, no.~6, pp. 66--76, Nov 2010.

\bibitem{whatmough2012vlsi}
P.~Whatmough, M.~Perrett, S.~Isam, and I.~Darwazeh, ``{VLSI} architecture for a
  reconfigurable spectrally efficient {FDM} baseband transmitter,'' \emph{IEEE
  Transactions on Circuits and Systems I: Regular Papers}, vol.~59, no.~5, pp.
  1107--1118, May 2012.

\bibitem{kim2011islanding}
J.-H. Kim, J.-G. Kim, Y.-H. Ji, Y.-C. Jung, and C.-Y. Won, ``An islanding
  detection method for a grid-connected system based on the {G}oertzel
  algorithm,'' \emph{IEEE Transactions on Power Electronics}, vol.~26, no.~4,
  pp. 1049--1055, Apr. 2011.

\bibitem{carugati2012variable}
I.~Carugati, S.~Maestri, P.~Donato, D.~Carrica, and M.~Benedetti, ``Variable
  sampling period filter {PLL} for distorted three-phase systems,'' \emph{Power
  Electronics, IEEE Transactions on}, vol.~27, no.~1, pp. 321--330, Jan. 2012.

\bibitem{wang1991pruning}
Z.~Wang, ``Pruning the fast discrete cosine transform,'' \emph{IEEE
  Transactions on Communications}, vol.~39, no.~5, pp. 640--643, May 1991.

\bibitem{skodras1994fast}
A.~Skodras, ``Fast discrete cosine transform pruning,'' \emph{IEEE Transactions
  on Signal Processing}, vol.~42, no.~7, pp. 1833--1837, Jul 1994.

\bibitem{Lecuire2012}
V.~Lecuire, L.~Makkaoui, and J.-M. Moureaux, ``Fast zonal {DCT} for energy
  conservation in wireless image sensor networks,'' \emph{Electronics Letters},
  vol.~48, no.~2, pp. 125--127, 2012.

\bibitem{kouadria2013low}
N.~Kouadria, N.~Doghmane, D.~Messadeg, and S.~Harize, ``Low complexity {DCT}
  for image compression in wireless visual sensor networks,'' \emph{Electronics
  Letters}, vol.~49, no.~24, pp. 1531--1532, 2013.

\bibitem{meher2014efficient}
P.~Meher, S.~Y. Park, B.~Mohanty, K.~S. Lim, and C.~Yeo, ``Efficient integer
  {DCT} architectures for {HEVC},'' \emph{Circuits and Systems for Video
  Technology, IEEE Transactions on}, vol.~24, no.~1, pp. 168--178, Jan. 2014.

\bibitem{Oppenheim2010}
A.~Oppenheim and R.~Schafer, \emph{Discrete-Time Signal Processing},
  3rd~ed.\hskip 1em plus 0.5em minus 0.4em\relax Pearson, 2010.

\bibitem{bc2012}
F.~M. Bayer and R.~J. Cintra, ``{DCT}-like transform for image compression
  requires 14 additions only,'' \emph{Electronics Letters}, vol.~48, no.~15,
  pp. 919--921, 19 2012.

\bibitem{Elliot1982}
D.~F. Elliot and K.~R. Rao, \emph{Fast Transforms: Algorithms, Analyses,
  Applications}.\hskip 1em plus 0.5em minus 0.4em\relax Academic Press, 1982.

\bibitem{Abadir2005}
K.~M. Abadir and J.~R. Magnus, \emph{Matrix Algebra}.\hskip 1em plus 0.5em
  minus 0.4em\relax Cambridge University Press, 2005.

\bibitem{bhaskaran1997}
V.~Bhaskaran and K.~Konstantinides, \emph{Image and Video Compression
  Standards}.\hskip 1em plus 0.5em minus 0.4em\relax Boston: Kluwer Academic
  Publishers, 1997.

\bibitem{USC_database}
\BIBentryALTinterwordspacing
{USC}-{SIPI} {I}mage {D}atabase. Signal and Image Processing Institute.
  University of Southern California. [Online]. Available:
  \url{http://sipi.use.edu/database/}
\BIBentrySTDinterwordspacing

\bibitem{penn1992}
W.~B. Pennebaker and J.~L. Mitchell, \emph{{JPEG} Still Image Data Compression
  Standard}.\hskip 1em plus 0.5em minus 0.4em\relax New York, NY: Van Nostrand
  Reinhold, 1992.

\bibitem{Wang2004}
Z.~Wang, A.~C. Bovik, H.~R. Sheikh, and E.~P. Simoncelli, ``Image quality
  assessment: from error visibility to structural similarity,'' \emph{IEEE
  Transactions on Image Processing}, vol.~13, no.~4, pp. 600--612, Apr. 2004.

\bibitem{sr-sim}
L.~Zhang and H.~Li, ``{SR-SIM}: A fast and high performance iqa index based on
  spectral residual,'' \emph{19th IEEE International Conference on Image
  Processing (ICIP)}, pp. 1473 -- 1476, 2012.

\bibitem{ehrgott2005multicriteria}
M.~Ehrgott, \emph{Multicriteria Optimization}, ser. Lecture Notes in Economics
  and Mathematical Systems.\hskip 1em plus 0.5em minus 0.4em\relax
  Springer-Verlag GmbH, 2005.

\bibitem{hm_software}
\BIBentryALTinterwordspacing
JCT-VC. (2015) {HM 10.0}. https://hevc.hhi.fraunhofer.de/. Joint Collaborative
  Team on Video Coding ({JCT-VC}). Fraunhofer Heinrich Hertz Institute.
  [Online]. Available: \url{https://hevc.hhi.fraunhofer.de/}
\BIBentrySTDinterwordspacing

\bibitem{xiph_database}
\BIBentryALTinterwordspacing
xiph.org. (2015) https://media.xiph.org/video/derf/. Xiph.org video test media.
  [Online]. Available: \url{https://media.xiph.org/video/derf/}
\BIBentrySTDinterwordspacing

\bibitem{Kimura2005}
N.~Kimura and S.~Latifi, ``A survey on data compression in wireless sensor
  networks,'' \emph{International Conference on Information Technology: Coding
  and Computing, ITCC}, vol.~2, pp. 8--13, 2005.

\bibitem{WMSN2007survey}
I.~F. Akyildiz, T.~Melodia, and K.~R. Chowdhury, ``A survey on wireless
  multimedia sensor networks,'' \emph{Computer Networks}, vol.~51, pp.
  921--960, 2007.

\end{thebibliography}
}

\end{document}